\documentclass[fleqn,usenatbib]{mnras}

\usepackage{newtxtext,newtxmath}

\usepackage[T1]{fontenc}

\DeclareRobustCommand{\VAN}[3]{#2}
\let\VANthebibliography\thebibliography
\def\thebibliography{\DeclareRobustCommand{\VAN}[3]{##3}\VANthebibliography}

\usepackage{graphicx}	
\usepackage{amsmath}	
\usepackage{multirow}
\usepackage{subcaption}

\title[Data-driven galaxy redshift distribution uncertainties]{Propagating data-driven galaxy redshift distribution uncertainties in 3$\times$2-pt analyses}

\author[Ruiz-Zapatero et al.]{
Jaime Ruiz-Zapatero,$^{1,2*}$ \thanks{$^*$jaime.ruiz-zapatero@ucl.ac.uk}
Qianjun Hang,$^{1}$
Yun-Hao Zhang,$^{3, 4}$
Benjamin Joachimi,$^1$
Joe Zuntz,$^{3}$
\newauthor 
Ian Harrison$^{5}$,
Carlos García-García$^{6, 7}$,
Alex Malz$^{8}$,
Benjamin Stölzner$^{9}$, 
\newauthor
and the LSST Dark Energy Science Collaboration
\\
$^1$ Department of Physics and Astronomy, University College London, Gower Street, London WC1E 6BT, UK \\
$^2$ Advanced Research Computing Centre, University College London, 90 High Holborn, London WC1V 6LJ, UK \\
$^3$ Institute for Astronomy, University of Edinburgh, Royal Observatory, Blackford Hill, Edinburgh, EH9 3HJ, UK \\
$^4$ Leiden Observatory, Leiden University, Einsteinweg 55, 2333 CC, Leiden, The Netherlands \\
$^5$ School of Physics and Astronomy, Cardiff University, The Parade, Cardiff, Wales CF24 3AA, UK \\
$^6$ Waterloo Centre for Astrophysics, University of Waterloo, Waterloo, ON N2L 3G1, Canada \\
$^7$ Department of Physics and Astronomy, University of Waterloo, Waterloo, ON N2L 3G1, Canada \\ 
$^8$ Space Telescope Science Institute, Baltimore, MD, USA \\
$^9$ Ruhr University Bochum, Faculty of Physics and Astronomy, Astronomical Institute (AIRUB), \\
German Centre for Cosmological Lensing, 44780 Bochum, Germany \\}
\date{Accepted XXX. Received YYY; in original form ZZZ}

\pubyear{\the\year{}}

\begin{document}
\label{firstpage}
\pagerange{\pageref{firstpage}--\pageref{lastpage}}
\maketitle

\begin{abstract}
Uncertainties in the radial distribution of galaxies, $\boldsymbol{n}(\boldsymbol{z})$, are one of the major contributions to the error budget of early Stage-IV galaxy survey analyses of weak gravitational lensing, galaxy clustering and galaxy-galaxy lensing (3$\times$2-pt).
Based on ensembles of simulated $\boldsymbol{n}(\boldsymbol{z})$ including stochastic and systematic variations, we study the impact of four different $\boldsymbol{n}(\boldsymbol{z})$ uncertainty models: shifts, shifts \& stretches, Gaussian processes (GP) and principal component analysis (PCA). 
Due to the high dimensionality of the latter models, we make use of state-of-the-art gradient-based inference methods as well as approximate analytical marginalisation schemes.
Our results show that Stage-IV 3$\times$2-pt analyses must go beyond simple shift \& stretch models. In particular, we advocate for the adoption of PCA models even in early Stage-IV surveys. Our results show that considering a five-parameters PCA model only degrades the constraint on the $S_{\rm 8}$ parameter by $5$ per cent with respect to the case when only a shift and a stretch parameter are included, while incurring half the bias in its constituents parameters, $\Omega_{\rm m}$ and $\sigma_{\rm 8}$.
We demonstrate that all models considered can be safely marginalised  analytically, with speed-ups of up to a factor of 25 depending on the dimensionality of the model. This will allow Stage-IV analyses to safely include higher-dimensional $\boldsymbol{n}(\boldsymbol{z})$ uncertainty models in their analysis at negligible additional computational cost.
\end{abstract}



\section{Introduction} \label{Sect: introduction}
Photometric galaxy surveys observe galaxies through a limited number of  filters as opposed to spectroscopic surveys where the full spectral energy distribution (SED) of the source's light is recorded \citep{First_photometry}. This incomplete measurement of the SED adds an extra source of uncertainty in the estimation of the source's redshift, leading to photometric redshifts being significantly less reliable than spectroscopic ones. Yet, how to characterise these uncertainties remains an open question. The literature is replete with methods to estimate the redshift distribution of  photometric galaxy samples such as template fitting \citep{BPZ}, weighted direct calibration given a sufficiently complete spectroscopic sample \citep{NIR_Nz1, NIR_Nz2}, clustering redshifts \citep{cross_corr_Nz1, cross_corr_Nz4, cross_corr_Nz2, cross_corr_Nz3} and shear ratios \citep{Kuijken_shear_ratio, DESY1_shear_ratios, DESY3_shear_ratios}. The problem of parameterising redshift distribution uncertainty is likely to continue to evolve with the characteristics of relevant surveys \citep{PhotometryFlavours, Photoz_challenge, photoz_review}.

In the context of projected two-point statistics, the value of the cosmological parameters is directly impacted by the radial distribution of the sample of galaxies as a whole, $n(z)$, rather than by the individual redshift probability distributions of galaxies, $p(z)$. Even when the redshift of each individual source in the sample has been estimated, deriving the corresponding $n(z)$ is non-trivial \citep{donotstack}. Moreover, characterizing the uncertainty in the $n(z)$ in a way that is representative of the uncertainty in the individual $p(z)$ is still an active area of research \citep{Yunhao1, Yunhao2}. Traditionally, $n(z)$ uncertainties were quantified a posteriori through a variety of arguments, often referring to simulations where the difference between the estimated and the ``true" galaxy distribution could be known \citep{DESY1-photo, HSCY1-photo, KIDSL-photo}. This approach puts significant constraints in the uncertainty models that can be calibrated robustly. Thus, the last generation of cosmological surveys, known as Stage-III, relied on simple redshift distribution uncertainty models where the mean and the width of the $n(z)$ was allowed to vary.

However, it is not clear that this methodology is able to encompass all the uncertainty in the $n(z)$ without introducing biases in the next generation of surveys, known Stage-IV surveys. For these reasons, modern galaxy surveys are moving towards data-driven ensemble methods to estimate the uncertainty of the $n(z)$ \citep{HSCY3-photo, DESY6-photo, mill_nz_bias,  RAIL}. This approach is based on generating an ensemble of possible $n(z)$ given the individual photometric redshift uncertainties, $p(z)$, of each source in the catalogue given a photometric redshift estimation code. Crucially, the resulting ensemble must account for both the statistical uncertainties, due to the individual $p(z)$, as well as systematic effects, due to uncertainties not represented in the catalogue such as the incompleteness of the sample. Once a representative ensemble has been derived, it can be reliably used to calibrate a wealth of $n(z)$ uncertainty models. 

In this work, we study the impact of different $n(z)$ uncertainty models on the cosmological constraints in a joint a analysis of the auto- and cross-correlations of the galaxy clustering and weak lensing fields), commonly known as a 3$\times$2-pt analysis, of mock Vera C. Rubin's observatory Legacy Survey of Space and Time Year 1 data, here onwards referred to as LSSTY1-like. We use the photometric redshift calibration of \citet{Yunhao1, Yunhao2} who provide an ensemble of $\boldsymbol{n}(\boldsymbol{z})$ distributions based on a combination of the \texttt{FlexZBoost}  \citep{Flexzboost} and \text{SOMPZ} \citep{SOM, 2021MNRAS.505.4249M} algorithms that encompasses both statistical and systematic uncertainties.  We leverage this ensemble of possible $n(z)$ distributions to calibrate a variety of  uncertainty models using the Dark Energy Science Collaboration (DESC) library, \texttt{nz\_prior}. We consider four different galaxy redshift distribution uncertainty models: a simple shift model, the most common shift and stretch model, a principal components decomposition and a Gaussian process.

Some of the models considered in this work significantly increase the dimensionality of the posterior being inferred to the point where traditional sampling techniques such as Metropolis Hastings \citep[][]{Metropolis} become inefficient. In order to tackle this problem, we propose two strategies, one numerical and one analytical. Numerically, we employ a Hamiltonian Monte Carlo \citep[HMC][]{HMC, Betancourt17} sampling scheme to marginalise over the additional dimensions. Analytically, we marginalise over the galaxy redshift distribution uncertainty parameters using the Laplace approximation \citep{Taylor&Kitcing} which Gaussianises the contribution of these parameters to the likelihood. Both techniques require computing the gradient of the likelihood efficiently. In order to do so, we write our likelihood using the auto-differentiable cosmological code \texttt{LimberJack.jl} \citep{LimberJack}.

The treatment of photometric redshift uncertainties has already received ample attention in the literature. The most relevant work to this paper is \citet{Ruiz_Zapatero_nzs} where a similar analysis to the one presented in this work was performed but limited to Weak Lensing (WL) analyses.  Before, \citet{Laplace_kiDS} studied the application to the Laplace approximation to the marginalisation of photometric uncertainties in Kilo-Degree Survey (KiDS) data. \citet{Boryana23} studied the validity of the Laplace approximation to marginalise over different nuisance parameters, including photometric uncertainties, in the Dark Energy Survey Year 1 (DESY1) 3$\times$2-pt analysis but limited its scope to a shifts \& stretches model. This work presents a novel thorough comparison of different redshift distribution uncertainty models for a 3$\times$2-pt Stage-IV like analysis as well as of how to marginalise over the parameters of those models. The outcomes of this comparison aim to provide useful insight into the analysis choices that collaborations such as the DESC \citep{LSST_SRD, lSST} and Euclid \citep{Euclid, Euclid_overview} will have to make when analysing their data.

This paper is structured as follows. In Section  \ref{Sect: Angular Power Spectra} we summarise the theory behind 3$\times$2-pt analyses. In Section \ref{Sect: data} we provide an overview the mock data vector. In Section \ref{Sect: Redshift Uncertainty} we show how galaxy redshift distribution uncertainty is characterised and incorporated through different models in our analysis. In Section \ref{Sect: Bayesian Inference} we discuss how we marginalise over the parameters of galaxy redshift distribution uncertainty models. In Section \ref{Sect: Results} we present the results of our analysis. Finally, in Section \ref{Sect: Conclusions} we summarise our findings and outline the takeaways for Stage-IV analyses.
 
\section{Angular Power Spectra} \label{Sect: Angular Power Spectra}
Assuming a flat, homogeneous and isotropic Universe, the angular power spectrum associated with the 2-pt correlation function of two projected tracers (labelled ${\rm f}$ and ${\rm g}$) of the matter density field can be computed as:
\begin{equation} \label{eq: Limber}
    C^{\rm fg}_\ell =\int \frac{d \chi}{\chi^2}q_{\rm f}(\chi)q_{\rm g}(\chi)P_{\rm fg}\left(\frac{\ell+1/2}{\chi},z\right) \, ,
\end{equation}
where $\ell$ refers to a given multipole, $z$ is the redshift, $\chi$ is the comoving distance, $q_f$ and $q_g$ are radial selection functions of the two fields being correlated and $P_{\rm fg}$ their three-dimensional power spectrum on the sphere, related to the underlying matter power spectrum. Note that Eq. \ref{eq: Limber} was computed under the Limber approximation \citep{Limber} which assumes that the kernels, $q_f$ and $q_g$, are slowly varying with respect the Bessel basis of projection onto the 2D sky which is valid $\ell$. In this  work we will focus on two particular tracers, galaxy clustering and weak gravitational lensing, which we will briefly describe below.

\subsection{Galaxy clustering}

Galaxy clustering (GC) carries information about the inhomogeneity of the underlying matter distribution. Galaxies form in the densest regions of the matter density field. Hence galaxies constitute a biased tracer of the matter density distribution. On large scales, we expect the bias to be linear \citep{Mo&White1996}. Thus we can establish the following relationship between the measured galaxy field and the underlying matter field:
\begin{equation} \label{eq: bg}
    \delta_{\rm g}(\boldsymbol{\theta}, z) \simeq b_{\rm g} \delta(\boldsymbol{\theta}, z) \, , 
\end{equation}
where $\delta(\boldsymbol{\theta}, z)$ and $\delta_g(\boldsymbol{\theta}, z)$ are the matter and galaxy overdensity fields, respectively, and $b_{\rm g}$ is a linear bias parameter. 

Incorporating this linear bias into the tracer kernel, the GC radial selection function is given by:
\begin{equation} \label{eq: clustering tracer}
    q_{\rm g}(\chi)=b_{\rm g} n(z) \frac{dz}{d\chi} \, ,
\end{equation}
where $n(z)$ is the normalised distribution of galaxies along the line of sight and $dz/d\chi$ relates the distributions of galaxies defined as a function of redshift to the tracer kernel defined as a function of radial comoving distance. 

\subsection{Weak lensing}
Weak gravitational lensing (WL) can be understood as a differentiable map between the true angle under which a transverse is observed in the sky and the perceived angle due to the presence of a lensing potential: 
\begin{equation} \label{eq: lensing_potential}
    \psi(\boldsymbol{\theta}, \chi) =  \int_0^\chi \frac{(\chi-\chi')}{\chi \chi'} \Phi(\boldsymbol{\theta}, \chi') d\chi'  \, ,
\end{equation}
where $\Phi(\boldsymbol{\theta}, \chi')$ is the gravitational potential of the matter distribution sourcing the lensing.

WL has two components: cosmic shear, related to the anisotropic distortion of the images, and the convergence field, which magnifies or reduces their size. Since the convergence field, $\kappa(\boldsymbol{\theta}, \chi)$, is related to the angular Laplacian of the lensing potential it can be directly related to the underlying matter density field through the Poisson equation leading to the following expression:
\begin{equation}
    \kappa(\boldsymbol{\theta}, \chi) = K_\ell \frac{3H_{\rm 0}^2 \Omega_{\rm m}}{2c^2} \int^\chi_0 \frac{d\chi'}{a(\chi') }\frac{\chi-\chi'}{\chi} \chi' \delta(\boldsymbol{\theta}, \chi) \,, 
\end{equation}
where $H_{\rm 0}$ is the Hubble parameter, $\Omega_{\rm m}$ is the cosmological matter density, $c$ is the speed of light and $a(\chi)$ is the scale factor and 
\begin{equation}
    K_\ell \equiv \frac{\ell (\ell+1)}{(\ell+1/2)^2} \, ,
\end{equation}
is a scale-dependent prefactor needed to transform  the 2D transverse angular Laplacian in the lensing equation to the full 3D angular Laplacian featured in the Newton-Poisson equation.  Therefore the associated radial selection function for the weak lensing tracer is given by 
\begin{equation} \label{eq: lensing_tracer}
  q_{\rm \kappa}(\chi) \equiv K_\ell \frac{3 H_0^2\Omega_{\rm{m}} \chi}{2 c^2 a(\chi)} \int_{z(\chi)}^\infty dz' n(z')\frac{\chi(z')-\chi}{\chi(z')} \, ,
\end{equation} 
where n(z) is once again the radial distribution of galaxies.

\section{Generating a LSSTY1-like 3x2-pt data vector} \label{Sect: data}

\begin{figure*}
  \includegraphics[width=1.0\textwidth]{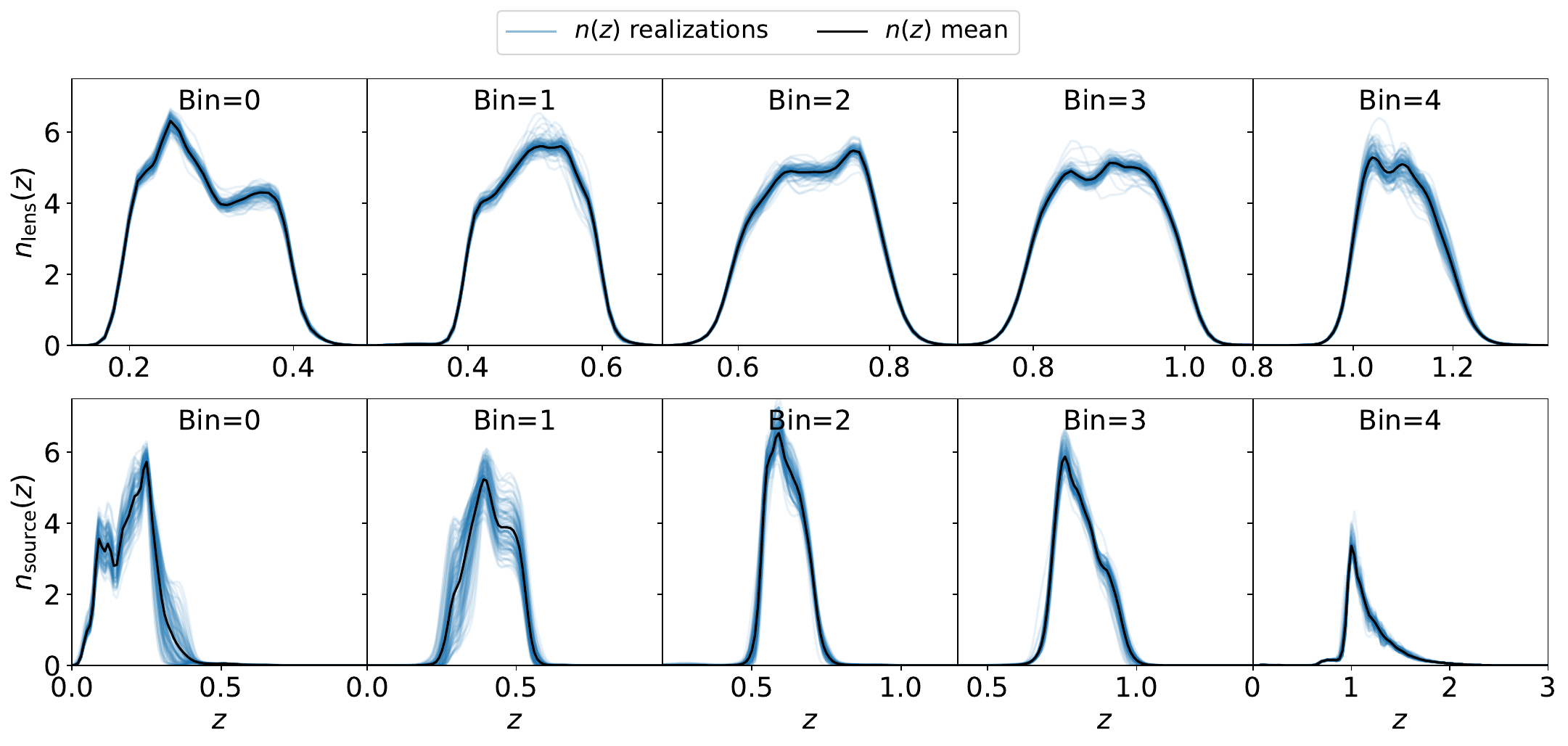} 
  \caption{The ensembles of redshift galaxy distributions from \citet{ Yunhao2} based on the CosmoDC2 catalogue augmented with the OpenUniverse24 catalogue for each tomographic bin for the lens (top row) and source samples (bottom row). Each ensemble contains the statistical uncertainties as given by the \texttt{FlexZBoost} algorithm as well as systematic uncertainties due to the incompleteness of the reference samples designed to match the excpect uncertainty of LSSTY1.}
  \label{fig:nzs}
\end{figure*}

\begin{table}  
     \caption{Fiducial values of cosmological and nuisance parameters used to generate the synthetic data vector of this analysis.}\label{Tab: best_cosmo}
      \centering
      \def\arraystretch{1.2}
      \begin{tabular}{|cc|cc|}
        \hline
        \multicolumn{4}{|c|}{\textbf{Fiducial Cosmology and Systematics}} \\
        \hline
        Parameter &  Value & Parameter &  Value\\  
        \hline 
        \multicolumn{2}{|c|}{\textbf{Cosmology}}    &      \multicolumn{2}{c|}{\textbf{Galaxy bias}}\\
        $\Omega_{\rm{m}}$  &  $0.27347$             &      $b_g^0$        & $0.87911$ \\
        $\Omega_{\rm{b}}$  &  $0.04217$             &      $b_g^1$        & $1.05894$ \\
        $h$                &  $0.71899$             &      $b_g^2$        & $1.22145$ \\
        $n_{\rm{s}}$       &  $0.99651$             &      $b_g^3$        & $1.35065$ \\
        $\sigma_{\rm{8}}$  &  $0.779007$            &      $b_g^4$        & $1.58909$ \\ \hline
                           &                        &      \multicolumn{2}{c|}{\textbf{Intrinsic Alignments}}\\ \hline
                           &                        &      $A_{\rm IA}$   & $0.251794$ \\
        \hline
      \end{tabular}
\end{table}

The joint analyses of the auto- and cross-correlation of galaxy clustering and weak lensing fields is commonly referred to as a 3$\times$2-pt analysis. Referring to the formalism laid on Sect. \ref{Sect: Angular Power Spectra}, this amounts to fitting $C_\ell^{\rm gg}$, $C_\ell^{\rm g\kappa}$ and $C_\ell^{\rm \kappa \kappa }$. Because of the different cuts and selections, 3$\times$2-pt analyses normally make use of different galaxy samples in their clustering and lensing fields. Thus it is common to refer to the clustering galaxies as the lenses sample, since they act as the foreground matter sourcing the lensing effect, and to the lensing sample as the sources, since they act at the background sources that get distorted. Moreover, these samples are further subdivided into redshift bins, known as tomographic bins, to allow for a more tailored modelling of systematics.

In order to create a LSSTY1-like 3$\times$2-pt data vector, we generate a noiseless synthetic data vector based on a $\boldsymbol{n}(\boldsymbol{z})$ calibration ensemble representative of the expected galaxy redshift distribution uncertainty in the LSSTY1 data release. This choice allows us to interpret the posteriors of different redshift uncertainty models more readily as well as spot any possible flaws in the methodology. Moreover, this ensures the self-consistency of the data products used in the analysis. 

In order to account for both statistical and systematic uncertainty in our radial distribution of galaxies, we adopt the redshift calibration of \citet{Yunhao2} based on the CosmoDC2 catalogue augmented with the OpenUniverse24 catalogue (See \citet{Yunhao1} for the details on the augmentation procedure).  The CosmoDC2 \citep{cosmoDC2} catalogue is based on a trillion-particle N-body simulation spanning  a volume of $4.225 \, \rm{Gpc}^3$. On the sky, it covers an area of $440 \, \rm{deg}^2$ up to a redshift of $z=3$ such that it matches the expected number density of LSSTY1 data. The CosmoDC2 catalogue is complete up to a magnitude depth of 28 in the LSST r-band. Galaxy properties such as the SED were assigned using a combination of the empirical model UniverseMachine \citep{2019MNRAS.488.3143B} tuned using the semi-analytical model Galacticus \citep{2012NewA...17..175B}. The realism of the generated catalogue, including its photometry, was validated up to LSST standards by \citet{2022OJAp....5E...1K}. The OpenUniverse24 simulations consist of a suite of realistic image simulations of a number of upcoming surveys, including LSST, covering $70 \, \rm{deg}^2$. OpenUniverse24 was built  as a direct successor to CosmoDC2, using the same halo catalogue and addressing some of the original shortcomings. Notably, OpenUniverse assigns galaxies properties using the newer \texttt{DiffSky} model which substantially improves the realism of optical and infrared photometry of objects \citep{2023MNRAS.518..562A, 2023MNRAS.521.1741H}. The OpenUniverse24 simulations were validated to LSST specifications by \citet{2025MNRAS.544.3799O}.

In their work, \citet{Yunhao2} developed a methodology to account for systematic effects in the photometric calibration of sources based on generating degraded spectroscopic-like datasets by simulating typical spectroscopic selection effects. To mitigate the biases and incompleteness in these degraded datasets, a Self Organising Map (SOM) based data augmentation strategy is then applied, enriching under-sampled regions in colour-magnitude-redshift space and producing more representative training samples. Then, photometric redshifts of each source were estimated using the \texttt{FlexZBoost} algorithm which employs gradient-boosted decision trees to estimate the redshifts of individual sources, $p(z)$, conditioned on its photometric properties. In this particular case, the algorithm was trained on combined degradation and augmentation datasets. Particularly, \citet{Yunhao2} employed the \texttt{FlexZBoost} implementation within the Redshift Assessment Infrastructure Layer, \texttt{RAIL}, available as part of the DESC open source software. The resulting ensembles of galaxy redshift distributions can be found in Fig. \ref{fig:nzs}. 

\begin{table}
\centering
\def\arraystretch{1.2}
\caption{ The scale cuts in multipole applied to each tomographic bin of the lens and sources samples \citep{Nicola_in_practice, Garcia-Garcia21}.} \label{Tab: scale_cuts}
\begin{tabular}{|p{2.5cm}|p{2.5cm}|}
\hline
 Lens  bin &  Scale cuts \\
 \hline 
 1 & $\ell < 145$ \\
 2 & $\ell < 225$ \\
 3 & $\ell < 298$ \\
 4 & $\ell < 371$ \\
 5 & $\ell < 435$ \\ 
 \hline
  Source  bin & Scale cuts \\ 
  \hline
  1-5 & $30 < \ell < 2000$ \\
\hline
\end{tabular}
\end{table}

We evaluate our theory code, \texttt{LimberJack.jl} \citep{LimberJack}, at the mean of the \citet{Yunhao2} photometric calibration (see black lines in Fig.~\ref{fig:nzs}) using the fiducial cosmology (see Table \ref{Tab: best_cosmo}).  In order to build the 3$\times$2-pt data vector, we consider the auto-correlation spectra of the lens sample bins, the auto- and cross-correlation spectra of the source bins as well as the cross-correlation spectra between the source and lens samples. In other words, we consider all possible correlations except the cross-correlations between the lens sample bins due to the relatively large impact of systematic effects compared to the cosmological signal in these correlations \citep{2021PhRvD.103d3503P}.  We then apply the scale cuts presented in Table \ref{Tab: scale_cuts}. Note that we apply particularly conservative scale cuts to the galaxy clustering auto-correlation spectra to exclude non-linear scales in the matter power spectrum beyond $k_{\rm max} = 0.15 \, {\rm Mpc}^{-1}$. This allows us to apply a simple linear bias model to include galaxy bias in our analysis \citep{Nicola_in_practice, Garcia-Garcia21}. We are also careful to exclude large scales where the Limber approximation used in Eq. \ref{eq: Limber}no longer holds \citep{NK5}. In addition to this, we also include intrinsic alignments as given by a simple NLA model \citep{BriddleLindsay, IA_rev} in terms of a single amplitude parameter.

For the covariance matrix, we rescale the \citet{Judit} Gaussian covariance matrix obtained for the analysis of the CosmoDC2 synthetic catalogue using the DESC code \texttt{TJPcov} \footnote{https://github.com/LSSTDESC/TJPCov}. However, the area covered by the CosmoDC2 catalog is much smaller than the expected $15000 \, \rm{deg}^2$ of LSSTY1. In order to address this discrepancy,  we multiply the original covariance by a factor $\alpha =  \rm{Area}_{\rm COSMOCDC2} / \rm{Area}_{\rm Y1} \approx 0.03$. This results in a tighter error budget that better approximates the expectation for LSST.

\section{\texttt{nz\_prior}: Data-Driven Redshift Uncertainties}  \label{Sect: Redshift Uncertainty}

In order to propagate the uncertainty in the galaxy distribution ensemble to cosmological constraints, we present \texttt{nz\_prior}\footnote{https://github.com/LSSTDESC/nz\_prior}. \texttt{nz\_prior} is a \texttt{Python} library of methods designed to calibrate priors for galaxy distribution uncertainty models based on a measured ensemble of samples of said galaxy distributions. Thus, it can run directly on the output of \texttt{RAIL} as the next stage of the LSST analysis pipeline. 

From here onwards we will assume the galaxy redshift distribution to be a vector variable, $\boldsymbol{n}(\boldsymbol{z})$, evaluated at an array of redshifts. Conceptually, the task boils down to finding the distribution of the parameters of the uncertainty model that produces an ensemble of $\boldsymbol{n}(\boldsymbol{z})$'s that matches the statistical properties of the original ensemble most closely. In other words, we want to perform dimensionality reduction that is as lossless as possible. In order to do so, \texttt{nz\_prior} projects each sample of the measured $\boldsymbol{n}(\boldsymbol{z})$ onto the space of the particular model:
\begin{equation}
    f: \mathbb{R}^{N_{\text{bins}}} \rightarrow \mathbb{R}^{N_{\alpha}}\, , 
\end{equation}
where $f$ is the mapping function and $N_{\rm bins}$ is the number of histogram bins of the $\boldsymbol{n}(\boldsymbol{z})$ in a given tomographic bin (i.e. the degrees of freedom of the original $\boldsymbol{n}(\boldsymbol{z})$) and  $N_{\rm \alpha}$ is the number of free parameters of the uncertainty model. Finally, \texttt{nz\_prior} uses the distributions of the uncertainty model parameters to calibrate a Gaussian prior. Thus, when a user employs \texttt{nz\_prior} to quantify and propagate galaxy redshift uncertainties, they make two assumptions:
\begin{itemize}
    \item the assumption of the chosen galaxy distribution uncertainty model (shifts, shifts \& stretches, etc...);
    \item the assumption that the distribution of model's parameters is Gaussian.
\end{itemize}
While the second assumption is rarely true except for the simplest models, we find that its impact on the statistical properties of the generated $\boldsymbol{n}(\boldsymbol{z})$ is below percentage level for all the models studied. On the other hand, the choice of model can lead to differences up to 25 per cent as we will discuss in Sect. \ref{Sect: Results}. Thus the first assumption is by far the lead contributor to the error budget. In what follows, we will cover the process of calibrating the parameters of the different uncertainty models. 

Throughout this section we will be taking two different forms of expectation values to calibrate the different $\boldsymbol{n}(\boldsymbol{z})$ uncertainty models. On the one hand, we want to compute expectation values over the calibration ensemble of $\boldsymbol{n}(\boldsymbol{z})$'s. We will denote the ensemble expectation values as $\langle  \rangle$. On the other hand, we are also interested in computing expectation values over redshift as given by a particular $\boldsymbol{n}(\boldsymbol{z})$. We will denote the redshift expectation value as $\{  \}$.

\subsection{Shifts and stretches}
Most cosmological analyses summarise the uncertainty in the calibration of the $\boldsymbol{n}(\boldsymbol{z})$ into two parameters. On the one hand, a shift parameter $\Delta z^\alpha$ that shifts the mean of the fiducial $\boldsymbol{n}(\boldsymbol{z})$. On the other hand, a stretch parameter that stretches the standard deviation of the $\boldsymbol{n}(\boldsymbol{z})$ distribution. This parametrization of the uncertainty is based on the assumption that the $\boldsymbol{n}(\boldsymbol{z})$ distribution is roughly Gaussian,
\begin{equation}
    \boldsymbol{n}(\boldsymbol{z}) \propto \exp[-(\boldsymbol{z}-\mu_{\rm z})^2/2\sigma_{\rm z}^2] \, ,
\end{equation}
where $\mu_{\rm z} = \{\boldsymbol{z}\}$ and $\sigma_{\rm z}^2 = \{\boldsymbol{z}^2\}-\{\boldsymbol{z}\}^2$. One can show that a new galaxy distribution $\tilde{\boldsymbol{n}}(\boldsymbol{z})$ with $\tilde{\mu}_{\rm z} = \mu_{\rm z} +\Delta_{\rm z}$ and $\tilde{\sigma}_{\rm z} = w_{\rm z}\sigma_{\rm z}$ can be generated by evaluating the original $\boldsymbol{n}(\boldsymbol{z})$ at:  
\begin{equation} \label{eq: z-tilde}
    \tilde{\boldsymbol{z}} = (\boldsymbol{z}-  \mu_{\rm z} - \Delta_{\rm z})/w_{\rm z} + \mu_{\rm z} \, .  
\end{equation}
Assuming an ensemble of $\boldsymbol{n}(\boldsymbol{z})$'s, it is possible to calibrate a prior for $(\Delta_{\rm z}, w_{\rm z})$ by solving the shift and stretch parameters of each $\boldsymbol{n}(\boldsymbol{z})$ realization in the ensemble with respect to a fiducial $\boldsymbol{n}(\boldsymbol{z})$. In our case, we will assume that the fiducial $\boldsymbol{n}(\boldsymbol{z})$ is the mean $\boldsymbol{n}(\boldsymbol{z})$ over the ensemble, $\overline{\boldsymbol{n}}(\boldsymbol{z}) = \langle \boldsymbol{n}(\boldsymbol{z})\rangle$. Thus for the $i$-th $\boldsymbol{n}(\boldsymbol{z})$ realization of the ensemble, the corresponding shift parameter is given by the difference between the mean $\boldsymbol{z}$ of the realization and the mean $\boldsymbol{z}$ of the mean $\boldsymbol{n}(\boldsymbol{z})$:
\begin{equation}
    \Delta_{\rm z}^{(i)} = \{\boldsymbol{z}\}^{(i)} - \{\langle \boldsymbol{z}\rangle\} \, .
\end{equation}
Similarly, the stretch parameter of the $i$-th realization is given by the ratio between the standard deviation of $\boldsymbol{z}$ of the $i$-th $\boldsymbol{n}(\boldsymbol{z})$ realization to the standard deviation of $\boldsymbol{z}$ of the mean $\boldsymbol{n}(\boldsymbol{z})$: 
\begin{equation}
    w_{\rm z}^{(i)} = \sqrt{\frac{\{\boldsymbol{z}^2\}^{(i)}-[\{\boldsymbol{z}\}^i]^2}{\{\langle \boldsymbol{z}^2 \rangle \}-\{\langle \boldsymbol{z} \rangle\}^2}} \, .
\end{equation}
Once the shift and stretch associated with each realization in the ensemble have been measured, a joint Gaussian prior for both parameters can be easily written down by measuring their mean and covariance over the ensemble of $\boldsymbol{n}(\boldsymbol{z})$'s.

It is important to note that both algorithms to find the shift and stretch parameters of a given $\boldsymbol{n}(\boldsymbol{z})$ ensemble assume that the $\boldsymbol{n}(\boldsymbol{z})$'s in the ensemble are roughly Gaussian. However, realistic radial galaxy distributions are often far from Gaussian which can lead so severe miss-calibrations. Given a ground truth, data vectors generated from these miss-calibrated priors often present a worse goodness-of-fit than simply fixing the photometric uncertainties.  In our testing we found the calibration of the shift parameter using the Gaussian assumption to be robust even when using significantly non-Gaussian redshift distributions, In other words including the shift parameter, even if misscalibrated, consistently improved the goodness of fit of the model. However, the stretch parameter proved to be much more sensitive to the Gaussian assumption. Our tests showed that miss-calibrated stretch priors often lead shifts \& stretches models performing worse than simpler shifts models.

In order to address this, we also include in \texttt{nz\_prior} an optimisation routine to calibrate the stretch parameter. In this routine, for every sample of the $\boldsymbol{n}(\boldsymbol{z})$ in the ensemble, we first find the corresponding shift using the Gaussian algorithm (see Eq. \ref{eq: z-tilde}). We then find the stretch parameter that reduces the distance between the sample $\boldsymbol{n}(\boldsymbol{z})$ and the modelled $\boldsymbol{n}(\boldsymbol{z})$ given the mean $\boldsymbol{n}(\boldsymbol{z})$ of the ensemble and the previously found shift parameter. We do so by writing the following distance metric:
\begin{equation}
\chi^{2} = (\boldsymbol{n}^{(i)}(\boldsymbol{z})-\langle \boldsymbol{n}(\tilde{\boldsymbol{z}}) \rangle) \textsf{C}_{\boldsymbol{n}(\boldsymbol{\boldsymbol{z}}), \boldsymbol{n}(\boldsymbol{z}')}^{-1} (\boldsymbol{n}^i(\boldsymbol{z})-\langle \boldsymbol{n}(\tilde{\boldsymbol{z}}) \rangle)^T \, , 
\end{equation}
where $\tilde{\boldsymbol{z}}$ is given by Eq.~(\ref{eq: z-tilde}) and $\textsf{C}_{\boldsymbol{n}(\boldsymbol{z}), \boldsymbol{n}(\boldsymbol{z}')}$ is the covariance of the ensemble of $\boldsymbol{n}(\boldsymbol{z})$ samples. This allows us to use scalar optimisation to find the optimal stretch parameter for each realization of the $\boldsymbol{n}(\boldsymbol{z})$ ensemble:
\begin{equation}
    w_{\rm z}^{(i)} = \rm{ArgMin}[\chi^{2}(w_{\rm z}^{(i)} | \boldsymbol{n}^{(i)}(\boldsymbol{z}),  \langle \boldsymbol{n}(\boldsymbol{z})\rangle, \Delta_{\rm z}^{(i)})] \, .
\end{equation}
Calibrating the joint Gaussian prior for the shift and stretch parameters is identical to using the Gaussian assumption. We find this methodology far more robust than the Gaussian approximation calibration for the typical shapes of galaxy redshift distributions.

\subsection{Principal components}
Since the measured redshift distribution of galaxies is defined by a process, i.e. a distribution over a family of functions, it is natural to try to capture its behaviour as an expansion in some given set of basis functions \citep{ACT_PCA, DES_PCA}. The fundamental challenge of this approach is to determine the best basis of functions to use. In other words, what basis of functions captures the most variability in the measured $\boldsymbol{n}(\boldsymbol{z})$ with the least possible number of expansion terms? Efforts are currently being made to find machine learning methods that can answer this question \citep{Yunhao2}. However, in this work we will opt for a classical answer to this problem: principal components.

Principal components analysis \citep[PCA, ][]{Jolliffe2002Principal, Bishop} answers two questions. First, what are the directions, $\boldsymbol{\phi}$, along which a covariance $\textsf{C}$ only results in a scaling (i.e. $\textsf{C} \boldsymbol{\phi} = \lambda \boldsymbol{\phi} $)? And, second, what is the magnitude of this scaling (i.e. $|\textsf{C} - \lambda 1| = 0$)? 

Thus, we can find the principal components by identifying the largest eigenvalues of the covariance matrix and finding their associated eigenvectors. Once obtained, the principal components define a family of orthogonal functions that can act as the basis of the expansion. Note that it is best practice to perform the PCA on the residuals of the $\boldsymbol{n}(\boldsymbol{z})$ ensemble, $\delta \boldsymbol{n}^{(i)}(\boldsymbol{z}) \equiv \boldsymbol{n}^{(i)}(\boldsymbol{z}) - \langle \boldsymbol{n}(\boldsymbol{z}) \rangle$, rather than on the $\boldsymbol{n}(\boldsymbol{z})$ itself such that the principal components do not have to capture large scale trends \citep{ACT_PCA}. Therefore, we can write a PCA model for the $\boldsymbol{n}(\boldsymbol{z})$ as:
\begin{equation} \label{eq: pca_expansion}
    \boldsymbol{n}(\boldsymbol{z})= \langle \boldsymbol{n}(\boldsymbol{z}) \rangle + \sum_k^{N_K} a_k \boldsymbol{\phi}_k(\boldsymbol{z}) \, , 
\end{equation}
where $\boldsymbol{\phi}_k$ denotes the $k$-th principal component of the residual ensemble of $\boldsymbol{n}(\boldsymbol{z})$'s and $a_k$ denotes the associated weight. 

Our goal now is to calibrate a prior on the expansion coefficients. In order to do so, we project the $\boldsymbol{n}(\boldsymbol{z})$ residuals on each eigenvector to obtain an ensemble of weights
\begin{equation}
    a^{(i)}_k = \int \delta \boldsymbol{n}^{(i)}(\boldsymbol{z}) \boldsymbol{\phi}_k(\boldsymbol{z}) d\boldsymbol{z} \, ,
\end{equation}
where $i$ denotes the sample in the ensemble and $k$ the expansion order. Similarly to the shift and stretch parameters, once the expansion weights of each $\boldsymbol{n}(\boldsymbol{z})$ have been measured, a joint Gaussian prior can be written down by measuring their mean and covariance. However, it is often useful to write $a$ in terms of a set of latent normally distributed parameters, $\boldsymbol{\alpha}$: 
\begin{equation}
    \boldsymbol{a} = \langle \boldsymbol{a} \rangle + \mathbb{L}[\boldsymbol{a}, \boldsymbol{a}']  \cdot \boldsymbol{\alpha}\,,
\end{equation}
where  $\boldsymbol{a} \in \mathbb{R}^K$ is the vector of expansion weights,  $\mathbb{L}[\boldsymbol{a}, \boldsymbol{a}'] $ refers to the lower Cholesky triangle of the covariance matrix of the expansion weights over the $\boldsymbol{n}(\boldsymbol{z})$ ensemble and $\boldsymbol{\alpha} \sim \mathcal{N}(0, 1) \in \mathbb{R}^K$. 
Now, since we are performing our expansion on the residuals of the $\boldsymbol{n}(\boldsymbol{z})$, the expectation value of the weights of the expansion should be $\langle \boldsymbol{a} \rangle \approx 0$. Similarly, an advantage of using eigenvectors as a basis for the expansion is that, by virtue of being orthogonal, the covariance of the weights is diagonal meaning that $\mathbb{L}[\boldsymbol{a}, \boldsymbol{a}'] \equiv \sigma_{\boldsymbol{a}} = \sqrt{\langle {\boldsymbol{a}}^2 \rangle-\langle \boldsymbol{a}\rangle^2}$. Plugging this back into Eq.~(\ref{eq: pca_expansion}) yields
\begin{equation}
    \boldsymbol{n}(\boldsymbol{z}) = \langle \boldsymbol{n}(\boldsymbol{z}) \rangle + \sum_k^K   \left[ \sigma_{a_k} \boldsymbol{\phi}_k(\boldsymbol{z}) \right]\alpha_k = \langle \boldsymbol{n}(\boldsymbol{z}) \rangle + \textsf{W}_{\rm PCA } \,   \boldsymbol{\alpha}  \, ,
\end{equation}
where $\textsf{W}_{\rm PCA}$ is a rectangular matrix where the $k$-th column is given by $\sigma_{a_k} \boldsymbol{\phi}_k(\boldsymbol{z})$.

\subsection{Gaussian Process}
Alternatively, one can characterise the measured ensemble of $\boldsymbol{n}(\boldsymbol{z})$ as a Gaussian Process (GP). This is equivalent to discarding all the information in the process beyond the second moments. The Gaussian approximation is generally a good approximation when the spread of the $\boldsymbol{n}(\boldsymbol{z})$ close to zero is small. This is because physically the $\boldsymbol{n}(\boldsymbol{z})$ cannot be negative leading to non-symmetrical (i.e. non-Gaussian) errors. However, if these errors are small compared to the errors in the bulk of the distribution, then the Gaussian approximation is valid \citep{HSCY3-photo}.

Similarly to when performing a principal component expansion, it is best practice to define the GP on the residuals of the $\boldsymbol{n}(\boldsymbol{z})$ \citep{Rasmussen}. Moreover, defining the GP on the fiducial redshift array of the $\boldsymbol{n}(\boldsymbol{z})$ might not always be desirable, especially when the array can have up to hundreds of entries which would lead to thousands of free parameters in the final analysis. Therefore, it is more efficient to introduce a secondary array of redshifts, $\boldsymbol{q}$, with reduced resolution, $Q$. 

Therefore, one can write a GP model for the redshift distribution of galaxies on the redshift array $\boldsymbol{q}$ as:
\begin{equation} \label{eq: gp_kernel}
    n(\boldsymbol{q}) = \langle n(\boldsymbol{q}) \rangle + \mathbb{L}[\delta n(\boldsymbol{q}),  \delta n(\boldsymbol{q}')] \, \boldsymbol{\alpha} \,  ,
\end{equation}
where $\boldsymbol{\alpha} \sim \mathcal{N}(0,1) \in \mathbb{R}^Q$ and $\delta n(\boldsymbol{q}) \equiv n(\boldsymbol{q}) - \langle n(\boldsymbol{q}) \rangle $. Note that in Eq.~(\ref{eq: gp_kernel}) the covariance of the calibration ensemble acts as the kernel of the GP. Typically, the kernel of a GP is a proposed smooth parametrization (radial kernel, sinusoidal kernel, ...) that is constrained a posteriori by the data. However, in our case we already have access to samples of the $\boldsymbol{n}(\boldsymbol{z})$ and can use them to set the noise properties of the GP without the need to assume a kernel. 

The final step is to map $n(\boldsymbol{q})$ back to $\boldsymbol{n}(\boldsymbol{z})$. Following the GP formalism, the posterior probability of a GP evaluated at an array $\boldsymbol{z}$ given an original array $q$, $P(\delta \boldsymbol{n}(\boldsymbol{z})|\delta n(\boldsymbol{q}))$, can be obtained by applying the following Wiener filter to $\delta n(\boldsymbol{q})$:
\begin{equation}
    \textsf{W}\left[\delta \boldsymbol{n}(\boldsymbol{z}), \delta n(\boldsymbol{q}) \right] = \textsf{C}\left[\delta \boldsymbol{n}(\boldsymbol{z}), \delta n(\boldsymbol{q}) \right] \left(\textsf{C}\left[\delta n(\boldsymbol{q}), \delta n(\boldsymbol{q}) \right] \right)^{-1} \, ,
\end{equation}
such that:
\begin{equation}
    \boldsymbol{n}(\boldsymbol{z}) = \langle \boldsymbol{n}(\boldsymbol{z}) \rangle + \textsf{W}\left[\delta \boldsymbol{n}(\boldsymbol{z}), \delta n(\boldsymbol{q}) \right] \mathbb{L}\left[\delta n(\boldsymbol{q}), \delta n(\boldsymbol{q}) \right]  \, \boldsymbol{\alpha} \, .
\end{equation}
Including the Cholesky decomposition of $\textsf{C}\left[(\delta n(\boldsymbol{q}), \delta n(\boldsymbol{q}) )\right]$ inside the Wiener filter we can simplify the previous expression as
\begin{gather}
    \boldsymbol{n}(\boldsymbol{z}) =  \langle \boldsymbol{n}(\boldsymbol{z}) \rangle + \textsf{C}\left[\delta \boldsymbol{n}(\boldsymbol{z}), \delta n(\boldsymbol{q}) \right] \left(\mathbb{L} \left[\delta n(\boldsymbol{q}), \delta n(\boldsymbol{q}) \right] \right)^{-1} \, \boldsymbol{\alpha} = \nonumber \\ = \langle \boldsymbol{n}(\boldsymbol{z}) \rangle + W_{\rm GP } \, \boldsymbol{\alpha} \,,
\end{gather}
where the columns of $W_{\rm{GP}}$ can be interpreted as the set of basis functions defined by the GP. Note that this is a remarkably similar result to the PCA model. This is because both the GP and PCA models are linear models defined on the residuals of the $\boldsymbol{n}(\boldsymbol{z})$ ensemble. The difference boils down to the different algorithms used to find the family of functions for the expansion. In the case of PCA, this is done by solving the characteristic equation of the covariance of the residuals. In the case of the GP, this is done by solving the Wiener filter of the same covariance.  

\section{Bayesian Inference}  \label{Sect: Bayesian Inference}
From here onwards, we will refer to the set of cosmological parameters as $\boldsymbol{\Omega}$ and to the nuisance parameters as $\boldsymbol{\nu}$ such that the total set of parameters is given by $\boldsymbol{\theta} = (\boldsymbol{\Omega}, \boldsymbol{\nu})$. In a Bayesian framework, the posterior distribution of $\boldsymbol{\theta}$ can be written as
\begin{equation} \label{eq: Bayes}
    P(\boldsymbol{\theta}|\boldsymbol{d}) = \frac{P(\boldsymbol{d}|\boldsymbol{\theta}) P(\boldsymbol{\theta})}{P(\boldsymbol{d})} \, ,
\end{equation}
where $P(\boldsymbol{d}|\boldsymbol{\theta})$ is the probability of the data given a model in terms of $\boldsymbol{\theta}$, $P(\boldsymbol{\theta})$ is the distribution of the parameters prior to observing the data (i.e. the distributions we derived in Section \ref{Sect: Redshift Uncertainty} for $\boldsymbol{\nu}$) and $P(\boldsymbol{d}) = \int P(\boldsymbol{d}|\boldsymbol{\theta}) P(\boldsymbol{\theta}) d\boldsymbol{\theta}$ is the probability of the data integrated over all the predictions of the model. Thus, the marginal distribution of $\Omega$ can be written as
\begin{equation} \label{eq: MargBayes}
    P(\boldsymbol{\Omega}|\boldsymbol{d}) = \int P(\boldsymbol{d} | \boldsymbol{\Omega}, \boldsymbol{\nu} ) P(\boldsymbol{\nu}) d\boldsymbol{\nu} \, .
\end{equation}

In this work we make use of the Bayesian inference framework \texttt{Turing.jl} \citep{ge2018t, Turing} to write our likelihood. \texttt{Turing.jl} is a probabilistic programming language (PPL) written in Julia. The main advantage of PPLs over traditional inference frameworks is that the posterior distribution is automatically derived from the specified priors and likelihood function allowing quick and easy iterations on the model. Moreover, being written in Julia,\texttt{Turing.jl} is fully auto-differentiable and is equipped with gradient-based samplers to tackle high-dimensional models. However, to make use of these gradient-based sampling techniques the theory computation inside the likelihood must also be auto-differentiable. Thus we compute our theory predictions using the same Julia auto-differentiable library used to compute synthetic data vector, \texttt{LimberJack.jl} \citep{LimberJack}. In the next sections we discuss the gradient based sampling algorithm used in this work in detail as well as the possibility of marginalising over certain parameter analytically.

\subsection{Sampling methods}
Computing  expectation values of high-dimensional distributions can be extremely numerically expensive. Traditionally, high-dimensional density integrals like Eq.~(\ref{eq: Bayes}) have been solved using Monte Carlo Markov Chain (MCMC) methods that stochastically propose samples for the numerical integration. Any MCMC method is defined by the transition kernel that proposes new samples given the current sample. Traditionally, MCMC methods have assumed that representative samples of the posterior live ``close" to one another in an isotropic way  \citep{Metropolis}. This simple assumption is generally true for low-dimensional distributions. However, it becomes progressively inaccurate as the dimensionality of the posterior increases due to the concentration-of-measure problem \citep{Betancourt17}.

Hamiltonian Monte Carlo (HMC) uses the gradient of the target distribution, in our case Eq.~(\ref{eq: MargBayes}), to move towards regions of high probability mass where representative samples live. In particular, HMC simulates Hamiltonian trajectories on the parameter space by introducing a set of auxiliary momentum variables, $p$. These are updated according to a force given by $-\nabla U(\boldsymbol{\theta})$ where $U(\boldsymbol{\theta}) = - \log P(\boldsymbol{\theta}|\boldsymbol{d})$. Thus, the total Hamiltonian of the dynamics is given by
\begin{gather}
    H(\boldsymbol{\theta}, \boldsymbol{p}) = K(\boldsymbol{\theta}, \boldsymbol{p}) + U(\boldsymbol{\theta})   = \frac{1}{2}(\boldsymbol{p} \Lambda^{-1}\boldsymbol{p}) - \log P(\boldsymbol{\theta}|\boldsymbol{d})\,,
\end{gather}
where $K(\boldsymbol{\theta}, \boldsymbol{p})$ acts as the kinetic contribution to the Hamiltonian. In standard HMC, the kinetic term is chosen to have a canonical form where $\Lambda$ is known as the mass matrix and acts as the proposal distribution for the momenta\footnote{See \citet{MCHMC} and \citet{MCHMC_2} for alternative kinetic terms}. 

For a sampled momenta, the dynamics of this Hamiltonian are evolved numerically during an arbitrary length using a symplectic integrator to preserve the volume of the posterior in terms of a finite step size, $\epsilon$. The position of the system at the last step becomes the new sample in the chain and the momenta are resampled. Finally, the sought-after posterior can be returned by marginalising the aforementioned Hamiltonian distribution over the momenta variables:
\begin{equation} \label{Eq:canonical_target}
    P(\boldsymbol{\theta}|\boldsymbol{d}) = \frac{1}{\mathcal{Z}} \int \exp(-H(\boldsymbol{\theta}, \boldsymbol{p})) \, d\boldsymbol{p} \, ,
\end{equation}
where $\mathcal{Z}$ is the canonical partition function given by:
\begin{equation} \label{eq: canonical_partition}
    \mathcal{Z} = \int \exp \left( \frac{-1}{2}{\boldsymbol{p}}^T\Lambda^{-1}{\boldsymbol{p}} \right) d {\boldsymbol{p}} \, .
\end{equation}

In this work we make use of the a self-tuning version of the HMC algorithm described above. The No U-Turns \citep[NUTS][]{NUTS} algorithm dynamically tunes the optimal values of the Hamiltonian dynamics step size and the mass matrix from which the momenta are proposed. Moreover, NUTS also introduces a dynamic criterion for when to stop evolving the Hamiltonian. NUTS identifies the point in the trajectory in which the momenta of the particle starts to turn, meaning that the system is starting to return to its original position. At this point, NUTS terminates the trajectory, giving name to the algorithm. This process ensures that the samples are as separated as possible.

\subsection{Analytical marginalisation} \label{sect: Laplace}

The Laplace approximation \citep{Kass1990TheVO, Hadzhiyska_2020, Laplace_kiDS, Boryana23, Ruiz_Zapatero_nzs} is a gradient-based analytical marginalisation scheme based on building a Gaussian approximation to the posterior distribution such that the integral in Eq.~(\ref{eq: MargBayes}) can be performed analytically. This is achieved by Taylor-expanding the log-posterior distribution around the best-fitting values of the dimensions we wish to Gaussianise. By virtue of expanding around the best-fitting values, the linear term of the expansion vanishes by definition, leaving only the zeroth term and the quadratic term. Since the zeroth term does not depend on $\boldsymbol{\nu}$, the dependence of the log-posterior on $\boldsymbol{\nu}$ becomes quadratic, leading to a Gaussian form for the posterior. 

In order to understand how this is done in practice, let us consider the best-fitting value of the nuisance parameters at fixed $\boldsymbol{\Omega}$. We define 
\begin{equation}
  \boldsymbol{\nu}_*(\boldsymbol{\Omega})\equiv {\rm arg\,\,max}_{\boldsymbol{\nu}} P(\boldsymbol{\Omega}, \boldsymbol{\nu} | \boldsymbol{d})\, ,
\end{equation}
such that $\boldsymbol{\nu}_*$ then satisfies
\begin{equation}\label{eq:minim}
  \left.\frac{\partial \chi^2}{\partial\boldsymbol{\nu}}\right|_{\boldsymbol{\nu}_*}=0,
\end{equation}
where we have defined $\chi^2 = -2\log P(\boldsymbol{\theta}|\boldsymbol{d})$. Following \cite{Taylor&Kitcing}, we can then approximate the distribution at each value of $\boldsymbol{\Omega}$ by expanding $\chi^2$ to second order in $\boldsymbol{\nu}$ around $\boldsymbol{\nu_*}$, i.e.:
\begin{equation} 
  \chi^2(\boldsymbol{\Omega},\nu)\simeq\chi^2_*(\boldsymbol{\Omega})+\Delta\boldsymbol{\nu}^T\textsf{F}_*\Delta\boldsymbol{\nu} \, ,
\end{equation}
where $\chi^2_*(\boldsymbol{\Omega})\equiv\chi^2(\boldsymbol{\Omega},\boldsymbol{\nu}_*)$, $\Delta\boldsymbol{\nu}\equiv\boldsymbol{\nu}-\boldsymbol{\nu}_*$, and $\textsf{F}_*$ is the Fisher information matrix:
\begin{equation}
  \textsf{F}_{*,ij}=\left.\frac{1}{2}\frac{\partial^2\chi^2}{\partial \nu_i\partial \nu_j}\right|_{\nu_*}.
\end{equation}

In this limit, the distribution is locally (i.e. at each $\boldsymbol{\Omega}$) a multivariate normal distribution in $\boldsymbol{\nu}$, and thus the integral in Eq.~\eqref{eq: MargBayes} can be solved analytically. The resulting marginalised posterior is given by
\begin{equation}\label{eq:laplace}
  \chi^2_{\rm m}(\boldsymbol{\Omega})\simeq\chi_*^2(\boldsymbol{\Omega})+\log\left\{\det\left[\textsf{F}_*(\boldsymbol{\Omega})\right]\right\}+{\rm const.} \, 
\end{equation}
From here onwards, we will refer to the first term in Eq.~(\ref{eq:laplace}) as the profile term (since it is equivalent to the profile likelihood) and to the second term as the Laplace term \footnote{It is worth noting that including the Laplace term in Eq.~\eqref{eq:laplace} should come at virtually no additional computational cost. Finding $\boldsymbol{\nu}_*(\boldsymbol{\Omega})$ requires solving for $\partial_{\nu}\chi^2=0$, which can be done efficiently using gradient descent methods. Finding the optimal step size in these algorithms often requires evaluating the Hessian of the function being minimised, and therefore the matrix $\mathcal{F}_*$ entering the Laplace term, is already a product of the minimisation algorithm. }.

However, the expression in Eq.~(\ref{eq:laplace}) can be further simplified to remove the need for optimising for $\boldsymbol{\nu_*}$ for each sampled set of $\boldsymbol{\Omega}$. When the likelihood of the data is Gaussian and the theory computation is linear on the nuisance parameters (or more generally, when the theory computation can be linearised on the nuisance parameters), \citep{Hadzhiyska_2020} showed that marginalising over said parameters is equivalent to adding an extra contribution to the covariance of the data. 

More formally, assume that the likelihood takes a Gaussian form:
\begin{equation}\label{eq:gauslike}
  \chi^2 =(\boldsymbol{d}-\boldsymbol{t})^T{\mathbb C}^{-1}(\boldsymbol{d}-\boldsymbol{t})+\chi^2_{\rm p\boldsymbol{\Omega}}(\boldsymbol{\Omega})+\chi^2_{\rm p\boldsymbol{\nu}}(\boldsymbol{\nu}).
\end{equation}
Here, $t(\boldsymbol{\Omega},\nu)$ is the theory vector, which depends on the model parameters, $\textsf{C}[\boldsymbol{d}, \boldsymbol{d}] \equiv \textsf{C}$ is the covariance matrix of the data, which we assume to be model-independent, and $\chi^2_{\rm p\boldsymbol{\Omega}}$ and $\chi^2_{\rm p\boldsymbol{\nu}}$ are the logarithm of the parameter priors. 

Now we linearise the theory prediction on the nuisance parameters, $\boldsymbol{t} \approx  \boldsymbol{t}_0 + J\nu$, where $\boldsymbol{t}_0$ is theory expectation evaluated at the expansion point and $J$ is the Jacobian of the theory with respect the nuisance parameters. Note that both $\boldsymbol{t}_0$ and $J$ are independent of $\nu$, but potentially dependent on $\boldsymbol{\Omega}$. However, \citet{Hadzhiyska_2020} showed that for narrow priors the dependency of the Jacobian on the cosmological parameters can be disregarded. Thus, in this study we keep $J$ fixed at the fiducial cosmology used to generate the data. In this limit, the $\chi^2$ is quadratic in $\boldsymbol{\nu}$ by construction, and the Laplace approximation is exact. Thus, the approximate relation between marginalisation and maximisation we outlined in the previous section becomes an equivalence when the data is Gaussian with a linear model in $\boldsymbol{\nu}$. Moreover, the profile term greatly simplifies to
\begin{equation}
  \chi^2_*={(\boldsymbol{d}-\boldsymbol{t})}^T\tilde{\textsf{C}}^{-1}{(\boldsymbol{d}-\boldsymbol{t})}\,,
  \label{eq:cov}
\end{equation}
where $\tilde{\mathbb C}$ is a modified covariance given by
\begin{equation}\label{eq:modcov}
  \tilde{\textsf{C}}={\textsf{C}}+\textsf{J}{\textsf{C}}_n \textsf{J}^T \, .
\end{equation}
obtained by simply assigning additional variance in quadrature to the modes of the data that align with the columns of $\textsf{J}$ (with this extra variance given by the $\nu$ parameter priors). We can then insert this modified covariance matrix into Eq.~(\ref{eq:gauslike}) to obtain:
\begin{equation} \label{eq:laplace_final}
  \chi^2_{\rm m}(\boldsymbol{\Omega}) \approx \,  (\boldsymbol{d}-\boldsymbol{t})^T\tilde{\textsf{C}}^{-1}(\boldsymbol{d}-\boldsymbol{t})+\chi^2_{\rm p\Omega}(\boldsymbol{\Omega}) + \, {\rm const.} \, . 
\end{equation}

To summarise, in the case of Gaussian data, negligible parameter dependence of the covariance matrix, and a theory model that is linear in the nuisance parameters, the Laplace approximation is exact. In this case there is a mathematical equivalence between marginalisation, $\chi^2$ minimisation, deprojection, and simply adding in quadrature the prior uncertainty on the marginalised parameters at the data level \citep{RybickiPress}. Following \citet{Ruiz_Zapatero_nzs}, who showed the validity of these assumptions to marginalise over galaxy redshift distribution uncertainty nuisance parameters in weak lensing analyses, we adopt Eq.~(\ref{eq:laplace_final}) to marginalise over the parameters of the galaxy redshift distribution uncertainty models described in Sect. \ref{Sect: Redshift Uncertainty}.

\begin{figure*} 
    \centering
    \includegraphics[width=\linewidth]{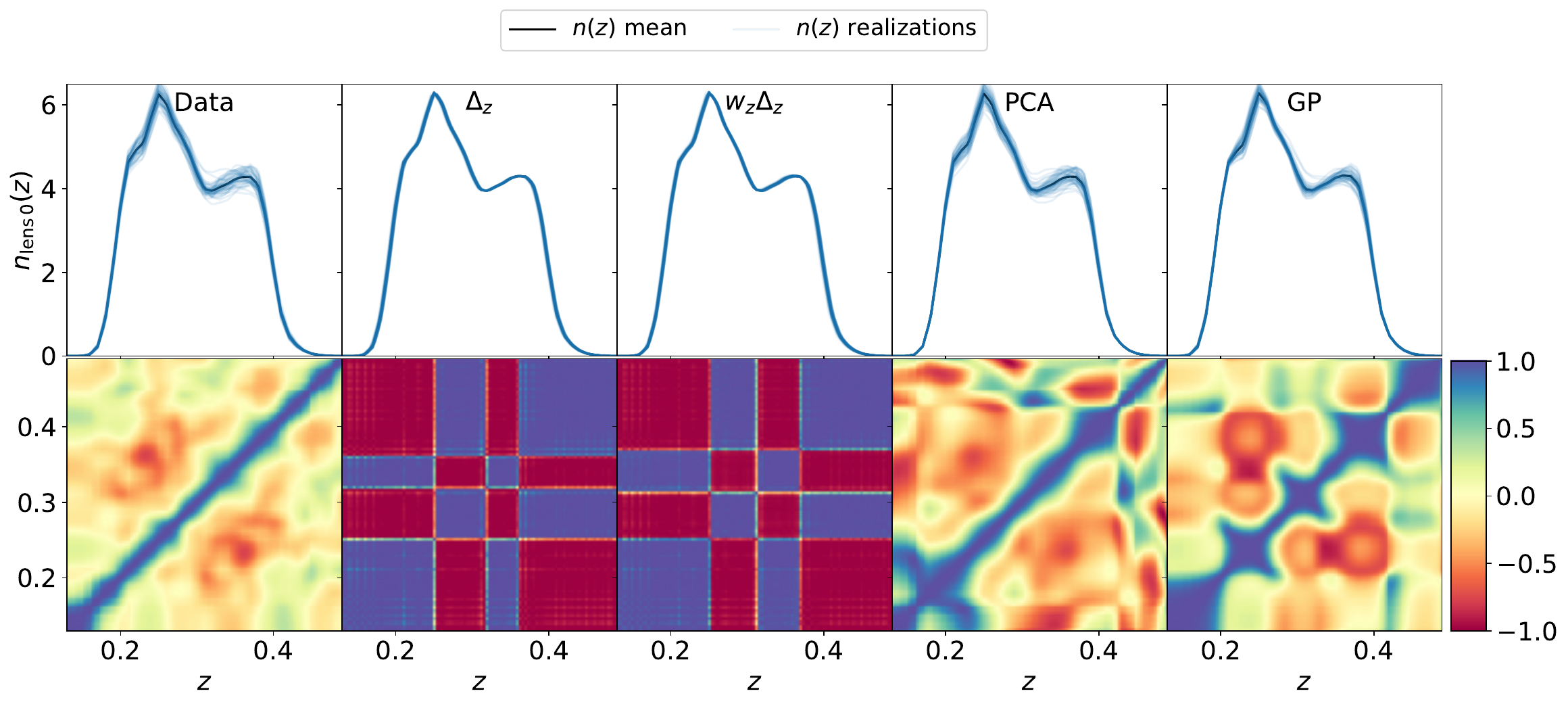}
    \caption{A comparison between the measured galaxy redshift distribution, $\boldsymbol{n}(\boldsymbol{z})$, for the first tomographic bin of the lens sample by \citet{Yunhao2} and the $\boldsymbol{n}(\boldsymbol{z})$ samples generated by different uncertainty models (shifts, shifts \& stretches, PCAs and GPs) after being calibrated on the former. The upper panels show overlapping samples of the processes. The lower panels show the correlation matrices of the processes.} \label{fig: nz_models}
\end{figure*}

\section{Results} \label{Sect: Results}

\subsection{Impact of n(z) uncertainty model choice}

The primary goal of this paper is to determine whether the choice of galaxy distribution uncertainty model has an impact in the posteriors of the cosmological parameters of interest. However, before we assess the impact of the different models on the cosmological parameters, we need to calibrate their priors first. For traditional models such as shifts and shifts \& stretches the number of parameters per tomographic bin is fixed by definition. However, it is not clear how many principal components or Gaussian process nodes should be included. In order to answer this question we study the ratio of the traces of the covariance matrix of the ensemble of $\boldsymbol{n}(\boldsymbol{z})$'s used for the calibration and the covariance matrix of the $\boldsymbol{n}(\boldsymbol{z})$ samples generated from the calibrated priors. Formally, we can write this as the $\zeta$-metric of a given model ``A":
\begin{equation} \label{eq: zeta-metric}
    \zeta^{\rm A} = \frac{\sum_i \textsf{C}^{\rm A}_{ii}}{\sum_i\textsf{C}^{\rm data}_{ii}}\;.
\end{equation}

Using Eq.~(\ref{eq: zeta-metric}), we observe that shifts capture approximately $50$ per cent  of the variance of the original ensemble of $\boldsymbol{n}(\boldsymbol{z})$'s. Adding a stretch parameter raises the variance captured to over $57$ per cent. Assigning 5 free parameters per tomographic bin for both the GP and PCA models manages to capture over $75$ per cent of the variance of the original ensemble while keeping the cost of the inference low. We investigated the impact of adding additional parameters to both models to obtain more representative priors.  We used Eq.~(\ref{eq:modcov}) to measure the approximate contribution of increasing the number of parameters on the likelihood. Pushing to 20 parameters per tomographic bin manages to capture up to $99$ per cent of the original variance. However, doing so only leads to a $1$ per cent increase in the impact on the likelihood with respect to when only 5 parameters per tomographic bins are included. Thus, we ultimately decided to include the reduced number of 5 parameters per tomographic bin for both PCA and GP models to keep the cost of the inference relatively low.

Having fixed the number of parameters of each model, we can now use \texttt{nz\_prior} to calibrate priors for the parameters of each model. In Fig. \ref{fig: nz_models} we show samples of the final calibrated models. In this figure we compare the $\boldsymbol{n}(\boldsymbol{z})$ ensemble measured \citet{Yunhao2} (top left panel) and its associated ensemble covariance (bottom left panel) with the equivalent quantities given by the different redshift uncertainty models (following columns). Visually, it is evident that different models result in different ensembles of $\boldsymbol{n}(\boldsymbol{z})$ even at prior level. The PCA and GP models are the only models able to capture the variations in the relative amplitudes in the two bumps of the $\boldsymbol{n(z)}$, a feature that has been shown to impact the corresponding angular power spectra \citep{Tessore&Harrison}. Moreover, the choice of uncertainty model not only impacts how much of the variance of the calibration ensemble is recovered but also the fidelity with which its correlation structure is captured. As expected the shifts and shifts \& stretches models induce very strong correlations between the bins of the histogram which looking fundamentally different to the diagonally dominated correlation structure of the data. PCA and GP models induce their own correlation structure. Nonetheless they resemble that of the data much more closely. Quantitatively, these discrepancies can be appreciated in the order of magnitude of the eigenvalues of the associated covariance matrices. Shifts and shifts \& stretches models can only recover the right order of magnitude for the first eigenvalue while PCA and GP model recover the right orders of magnitude for up to the fifth eigenvalue. Thus we observe a clear dimensionality scaling between the number of parameters of the model and fidelity with which the correlation structure of the calibration ensemble is matched.

To propagate these differences to the posterior distribution we run MCMC's using the NUTS algorithm for each of the models as described in Section \ref{Sect: Bayesian Inference}, namely: shifts, shifts \& stretches, PCAs and GPs.  We combine the derived priors for the different $\boldsymbol{n}(\boldsymbol{z})$ uncertainty models with prior distributions for the cosmological parameters as well as for the different systematic effects. The specific priors on each parameter can be found in Table \ref{Tab: cosmo_priors}. Additionally, we run an analysis where the photometric redshift uncertainties were kept fixed to establish a baseline for the different methods.

\begin{table}
      \caption{Prior distributions for the cosmological and nuisance parameters of this analysis.  $\mathcal{N}(a, b)$ denotes a Gaussian distribution with mean $a$ and standard deviation $b$. $U(c, d)$ denotes a uniform distribution with a lower bound $c$ and an upper bound $d$. Hence, $U(\mathcal{N}(a, b), c, d)$ denotes a truncated Gaussian distribution.} \label{Tab: cosmo_priors}
      \centering
      \def\arraystretch{1.2}
      \begin{tabular}{|cc|cc|}
              \hline
        \multicolumn{4}{|c|}{\textbf{Cosmology and systematics priors}} \\
        \hline
        Parameter &  Prior & Parameter &  Prior\\  
        \hline 
        \multicolumn{2}{|c|}{\textbf{Cosmology}}               &      \multicolumn{2}{c|}{\textbf{Galaxy bias }}\\
        $\Omega_{\rm{m}}$  &  $U (0.2, 0.6)$                   &      $b_g^i $       & $U (0.5, 2.5)$ \\
        $\Omega_{\rm{b}}$  &  $U (0.028, 0.065)$               &      \multicolumn{2}{c|}{\textbf{Intrinsic Alignments}}\\
        $h$                &  $U (\mathcal{N}(0.72, 0.05), 0.64, 0.82)$       &      $A_\mathrm{IA,0} $ & $U(-1.0, 1.0)$\\
        $n_{\rm{s}}$       &  $U (0.84, 1.1)$                  &      &  \\
        $\sigma_{\rm{8}}$  &  $U(0.4, 1.2)$                    &      & \\
        \multicolumn{2}{|c|}{\textbf{Shifts}}                     &  \multicolumn{2}{c|}{\textbf{Stretches}}\\
        $\Delta z^{\rm lens \, 1}$  &  $\mathcal{N}(0, 0.049)$    &  $w_{\rm z}^{\rm lens \, 1}$  &  $\mathcal{N}(1, 0.067)$ \\
        $\Delta z^{\rm lens \, 2}$  &  $\mathcal{N}(0, 0.043)$    &  $w_{\rm z}^{\rm lens \, 2}$  &  $\mathcal{N}(1, 0.109)$ \\
        $\Delta z^{\rm lens \, 3}$  &  $\mathcal{N}(0, 0.036)$    &  $w_{\rm z}^{\rm lens \, 3}$  &  $\mathcal{N}(1, 0.042)$ \\
        $\Delta z^{\rm lens \, 4}$  &  $\mathcal{N}(0, 0.034)$    &  $w_{\rm z}^{\rm lens \, 4}$  &  $\mathcal{N}(1, 0.031)$ \\ 
        $\Delta z^{\rm lens \, 5}$  &  $\mathcal{N}(0, 0.031)$    &  $w_{\rm z}^{\rm lens \, 5}$  &  $\mathcal{N}(1, 0.053)$ \\ 
        $\Delta z^{\rm source \, 1}$  &  $\mathcal{N}(0, 0.073)$  &  $w_{\rm z}^{\rm source \, 1}$  &  $\mathcal{N}(1, 0.092)$ \\ 
        $\Delta z^{\rm source \, 2}$  &  $\mathcal{N}(0, 0.068)$  &  $w_{\rm z}^{\rm source \, 2}$  &  $\mathcal{N}(1, 0.078)$ \\ 
        $\Delta z^{\rm source \, 3}$  &  $\mathcal{N}(0, 0.054)$  &  $w_{\rm z}^{\rm source \, 3}$  &  $\mathcal{N}(1, 0.18)$ \\ 
        $\Delta z^{\rm source \, 4}$  &  $\mathcal{N}(0, 0.052)$  &  $w_{\rm z}^{\rm source \, 4}$  &  $\mathcal{N}(1, 0.05)$ \\ 
        $\Delta z^{\rm source \, 5}$  &  $\mathcal{N}(0, 0.124)$  &  $w_{\rm z}^{\rm source \, 5}$  &  $\mathcal{N}(1, 0.072)$ \\ 
        \multicolumn{2}{|c|}{\textbf{PCAs}}                     &  \multicolumn{2}{c|}{\textbf{GPs}}\\
        $\alpha_{\rm{PCA} \, k}^{\rm lens \, i}$  &  $\mathcal{N}(0, 1)$   &  $\alpha_{\rm{GP} \, k}^{\rm lens \, i}$  &  $\mathcal{N}(0, 1)$ \\
        $\alpha_{\rm{PCA} \, k}^{\rm source \, i}$  &  $\mathcal{N}(0, 1)$ & $\alpha_{\rm{GP} \, k}^{\rm source \, i}$  &  $\mathcal{N}(0, 1)$ \\
        \hline
      \end{tabular}
\end{table}

\begin{figure*}
    \centering
    \includegraphics[width=\linewidth]{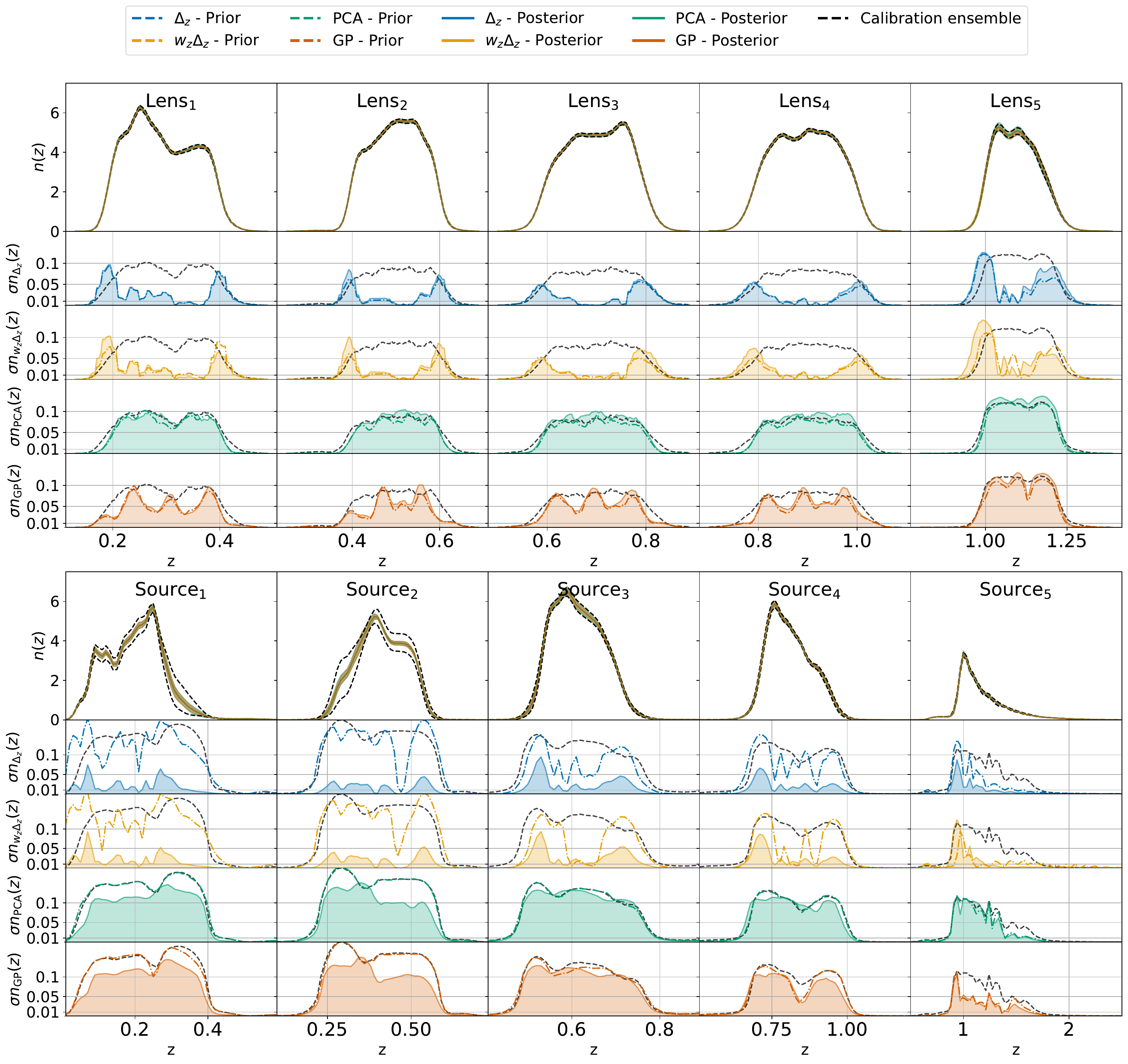}
    \caption{The posteriors for the galaxy distribution of each tomographic bin given by different uncertainty models. Upper panels show the lens sample bins. Lower panels show the source sample bins. The first row of panels in each block shows a direct comparison between the galaxy distributions obtained for each model. In the rows below we show the standard deviation on the $\boldsymbol{n}(\boldsymbol{z})$ posterior obtained by every model considered at every redshift (solid coloured contours). In order to understand whether the constraints are prior-dominated or not, we also overplot the standard deviation of the prior calibrated for each model (dashed coloured contours). Finally, we also include the standard deviation of the original ensemble of $\boldsymbol{n}(\boldsymbol{z})$'s used to calibrate each of the priors for reference  (black dashed lines). Blue contours correspond to the shifts ($\Delta_{\rm z}$) model posterior, yellow contours correspond to the shifts \& stretches ($\Delta_{\rm z} w_{\rm z}$) model posterior, green contours correspond the PCA model posterior and orange contours correspond to the GP model posterior.} \label{fig: nz_post}
\end{figure*}

\begin{figure} 
    \centering
    \includegraphics[width=0.95\linewidth]{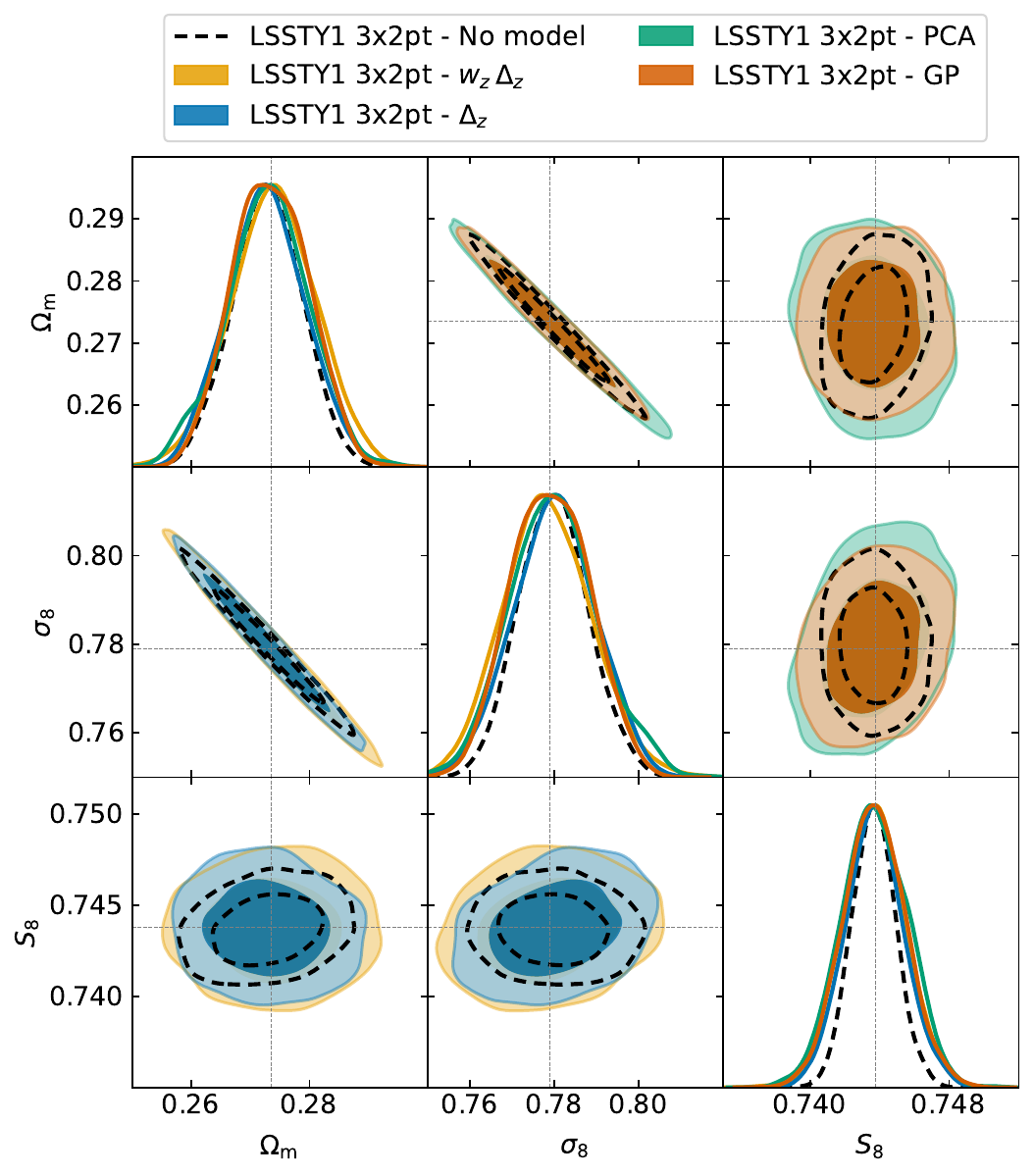}
    \caption{The 1 \& 2D marginalised constraints for the cosmological parameters $\Omega_{\rm m}$, $\sigma_{\rm 8}$ and $S_{\rm 8}$, accounting for galaxy distribution uncertainties numerically using different models. Markers show the cosmology used to generate the data. Black dashed contours correspond to when no model was considered (i.e. galaxy distribution uncertainties were not considered in the analysis).  Blue contours correspond to the shifts ($\Delta_{\rm z}$) model posterior, yellow contours correspond to the shifts \& stretches ($\Delta_{\rm z} w_{\rm z}$) model posterior, green contours correspond the PCA model posterior and orange contours correspond to the GP model posterior.} \label{fig: model_comp}
\end{figure}

\begin{figure} 
    \centering
    \includegraphics[width=0.95\linewidth]{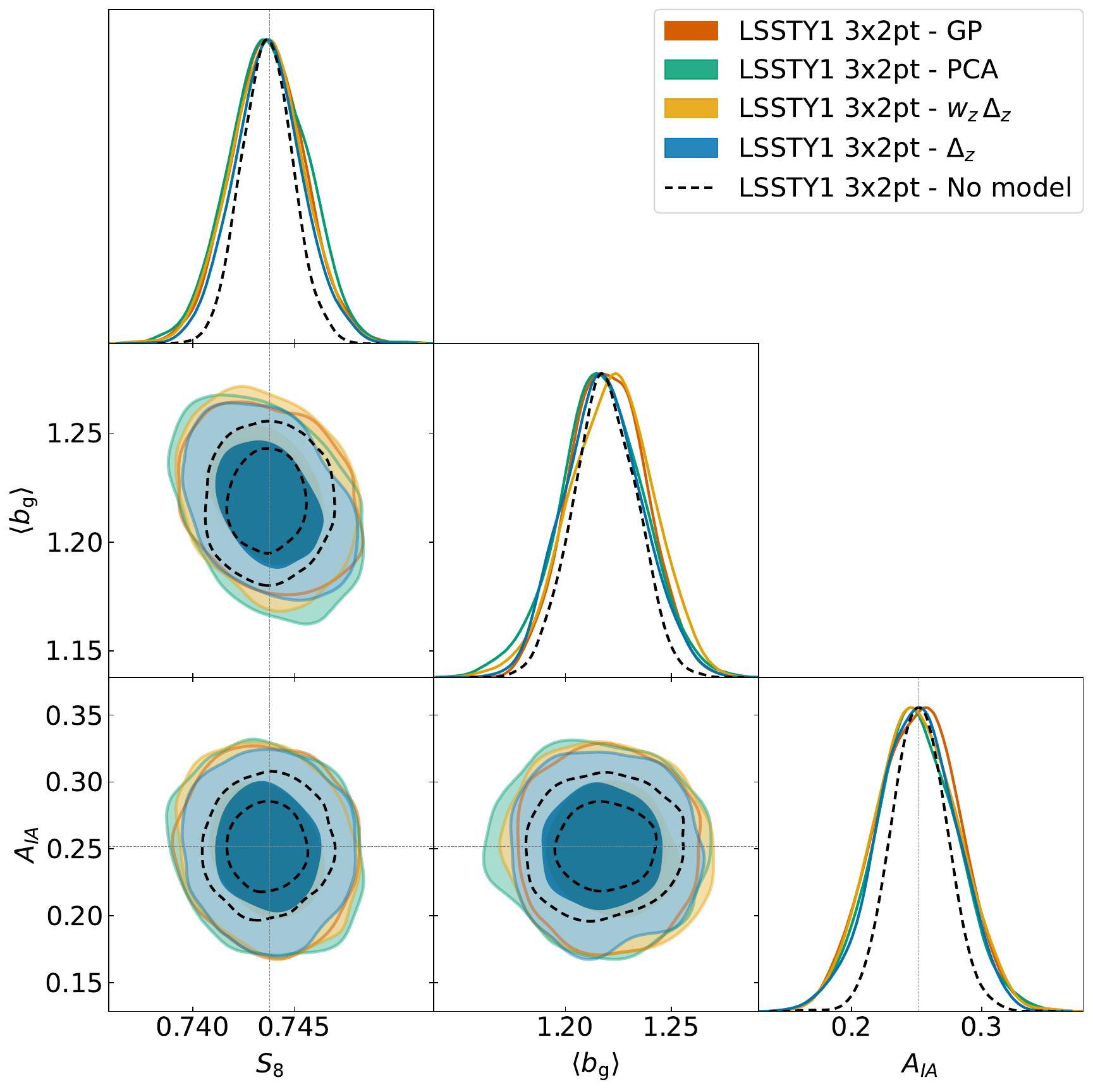}
    \caption{The 1 \& 2D marginalised constraints for the cosmological parameters $S_{\rm 8}$ and $\langle b_g\rangle$, the average galaxy bias across all tomographic bins, accounting for galaxy distribution uncertainties using different models. Black dashed contours correspond to when no model was considered (i.e. galaxy distribution uncertainties were not considered in the analysis).  Blue contours correspond to the shifts ($\Delta_{\rm z}$) model posterior, yellow contours correspond to the shifts \& stretches ($\Delta_{\rm z} w_{\rm z}$) model posterior, green contours correspond the PCA model posterior and orange contours correspond to the GP model posterior.} \label{fig: bias}
\end{figure}

\begin{figure} 
    \centering
    \includegraphics[width=0.85\linewidth]{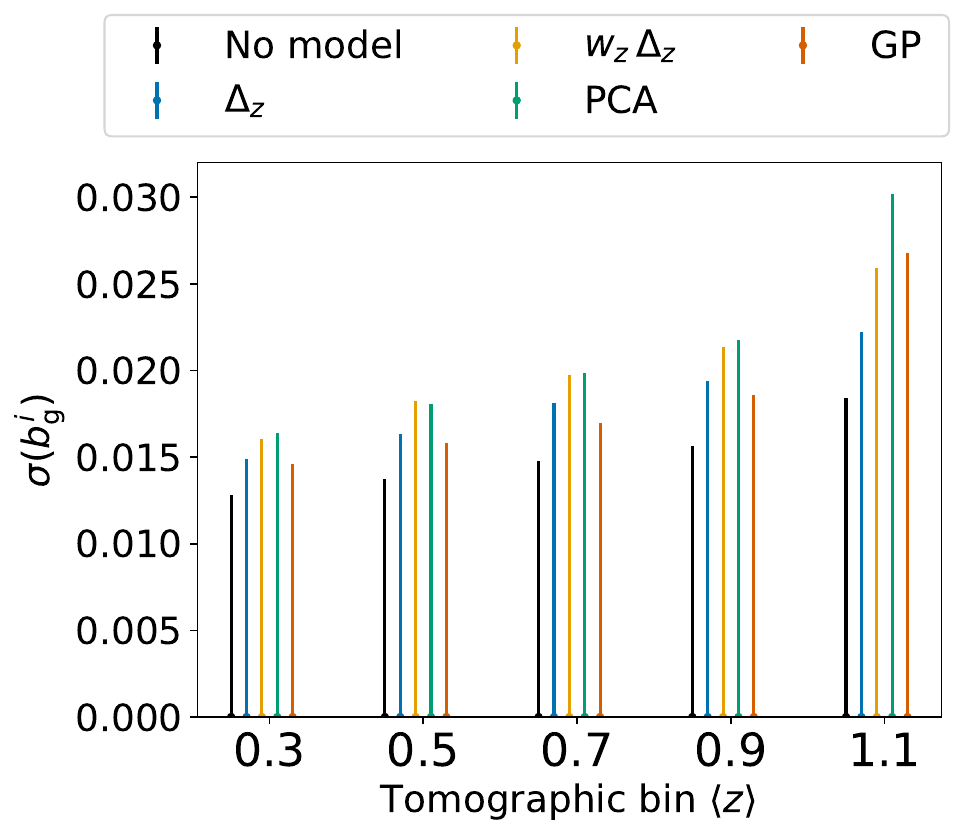}
    \caption{The standard deviation of the galaxy bias parameter of each tomographic bin in the lens sample. The labels on the horizontal axis correspond to the mean redshift of each tomographic bin. The different error bars represent the constraints from the different galaxy distribution uncertainty models considered in this work. Black lines correspond to when no model was considered (i.e. galaxy distribution uncertainties were not considered in the analysis).  Blue lines correspond to the shifts ($\Delta_{\rm z}$) model posterior, yellow lines correspond to the shifts \& stretches ($\Delta_{\rm z} w_{\rm z}$) model posterior, green lines correspond the PCA model posterior and orange lines correspond to the GP model posterior.} \label{fig: bias_z}
\end{figure}

\begin{figure} 
    \centering
    \includegraphics[width=0.85\linewidth]{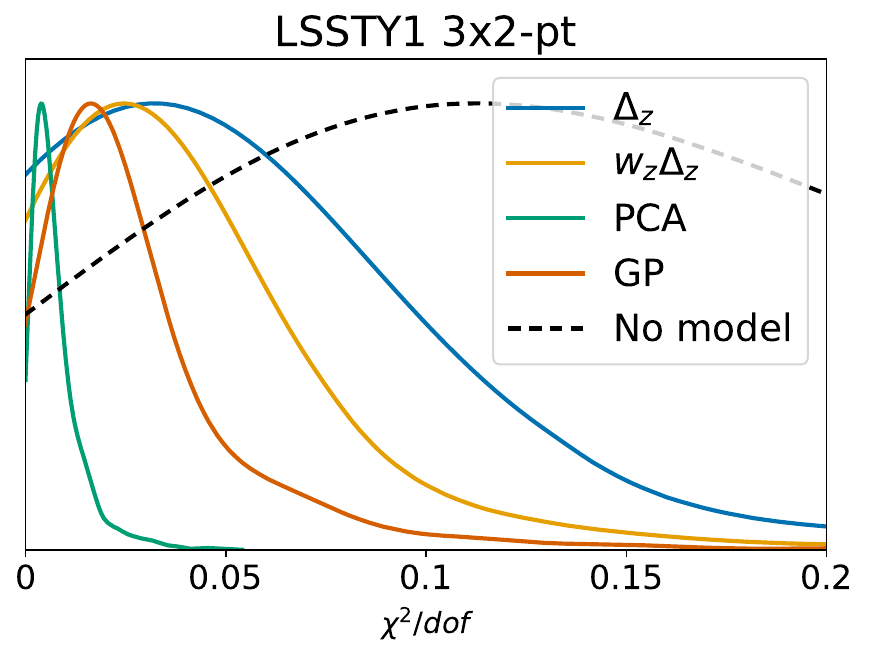}
    \caption{Reduced $\chi^2$ distributions obtained when assuming a given galaxy distribution uncertainty model to fit noiseless data vectors based on the samples in the $\boldsymbol{n}(\boldsymbol{z})$ calibration ensemble for shared cosmology. Thus the $\chi^2$ distributions shown represent the error incurred by assuming a given model. Models with a higher error will lead to a higher bias in the cosmology parameters. The black dashed line corresponds to when no model was considered (i.e. galaxy distribution uncertainties were not considered in the analysis). The blue line corresponds to the shifts ($\Delta_{\rm z}$) model posterior, the yellow line corresponds to the shifts \& stretches ($\Delta_{\rm z} w_{\rm z}$) model posterior, the green line corresponds the PCA model posterior and the orange line correspond to the GP model posterior} \label{fig: xi2_bias}
\end{figure}

\begin{figure} 
    \centering
    \includegraphics[width=0.85\linewidth]{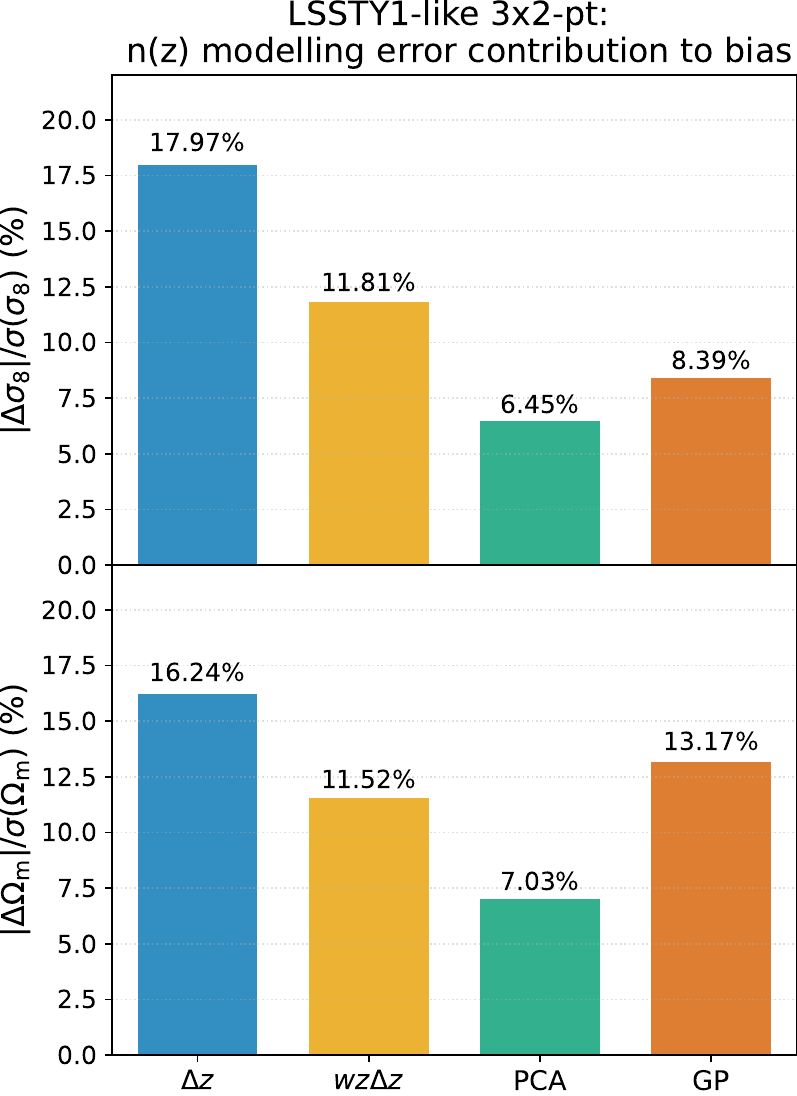}
    \caption{Square root bias over standard deviation as a percentage induced on the cosmological parameters $\Omega_{\rm m}$ (bottom panel) and $\sigma_{\rm 8}$ (top panel) purely due to the choice of $\boldsymbol{n}(\boldsymbol{z})$ uncertainty model. The bias measurement was obtained using a linear approximation to fit theory vectors generated using $\boldsymbol{n}(\boldsymbol{z})$ samples from the calibration ensemble using theory vectors generated based on the $\boldsymbol{n}(\boldsymbol{z})$ given by each model. The standard deviation was obtained running Monte Carlo chains for each model. We observe that the PCA model halves the bias in both cosmological parameters with respect to the shifts \& stretches model.}
    \label{fig: bias_variance}
\end{figure}

In Fig.~\ref{fig: nz_post} we compare the posteriors for the galaxy redshift distribution of each tomographic bin given by different uncertainty models. Upper (lower) panels show the lens (source) sample bins. Black dashed lines show the spread of the ensemble of $\boldsymbol{n}(\boldsymbol{z})$'s from which the priors of each model were calibrated (see Section~\ref{Sect: Redshift Uncertainty}). Note that, due to the strict requirements of LSSTY1 on the calibration, it is difficult to tell the difference between the posteriors of the different models. Thus, we include a subpanel for each panel where we plot the residuals of each $\boldsymbol{n}(\boldsymbol{z})$ posterior.

We observe that 3$\times$2-pt analyses are capable of significantly constraining the $\boldsymbol{n}(\boldsymbol{z})$ at a process level, resulting in qualitatively different posteriors for each of the different galaxy redshift distribution uncertainty models. Given the precision requirements of Stage-IV surveys, these differences are only visible by looking at the residuals of each tomographic bin. Visually, we can observe that the higher-dimensional models lead to broader processes than the lower-dimensional ones. In the case of the lens samples we observe that constraints on the $\boldsymbol{n}(\boldsymbol{z})$ for the shifts and shifts \& stretches models are prior-driven. However, in the case of the source sample the self-calibration of the data significantly reduces the spread of the $\boldsymbol{n}(\boldsymbol{z})$'s several-fold. Moreover, we can see that for both lenses and sources the GP and PCA models constraints are prior-dominated. This means that the data cannot self-calibrate the higher-dimensional models. Since the PCA and GP priors capture a much larger portion of the variance in the calibration ensemble of $\boldsymbol{n}(\boldsymbol{z})$'s, the resulting posteriors are 2 to 5 times wider than those of the shifts and shifts \& stretches models.

Having shown that the different galaxy redshift distribution uncertainty models lead to different $\boldsymbol{n}(\boldsymbol{z})$ posteriors, we can now ask the question whether these differences lead to different cosmological parameters posteriors. We focus on the cosmological parameters $\Omega_{\rm m}$, the cosmological matter density, and $\sigma_{\rm 8}$, the variance of the matter field in spheres of 8 $\texttt{Mpc}/h$ radius. Moreover, we will investigate the derived parameter $S_{\rm 8} = \sigma_{\rm 8} \sqrt{\Omega_{\rm m}/0.3}$, which is constructed to measured the stretch of the degeneracy between $\Omega_{\rm m}$ and $\sigma_{\rm 8}$ in weak lensing analyses. In Fig. \ref{fig: model_comp} we plot the 1D \& 2D marginalised constraints for the cosmological parameters $\Omega_{\rm m}$, $\sigma_{\rm 8}$ and $S_{\rm 8}$ resulting from the previously described chains.  Our results show that including a shift parameter to account for redshift uncertainties in 3$\times$2-pt analyses leads to a $12$ per cent and $18$ per cent degradation in the $\Omega_{\rm m}$ and $\sigma_{\rm 8}$ constraints, respectively. In combination, this leads to a nearly $34$ per cent loss in constraining power in the parameter $S_{\rm 8}$. Including an additional stretch parameter widens the $\Omega_{\rm m}$ and $\sigma_{\rm 8}$ constraints by  $23$ per cent and $28$ per cent respectively with respect when no model is considered. In combination, considering a shifts \&  stretches model degrades the $S_8$ constraint by over $42$ per cent with respect to when no model is considered or $5$ per cent with respect just considering a shift parameter. Moreover, we note that the constraints on all the stretch parameters are not prior-dominated, meaning that the difference in constraints cannot be due to the stretch prior being too conservative or miss-calibrated.

Considering higher-dimensional models such as GP and PCA leads to a further degradation of the constraint. In terms of $\Omega_{\rm m}$ and $\sigma_{\rm 8}$, the PCA model degrades the constraints by  $20$ per cent and $29$ per cent respectively with respect the case where no uncertainty model was considered. This leads to a $51$ per cent wider $S_{\rm 8}$ constraint. The GP model leads to a smaller degradation finding a $8$ per cent wider $\Omega_{\rm m}$, a $13$ per cent wider $\sigma_{\rm 8}$ and $43$ per cent wider $S_{\rm 8}$ constraint with respect the case where no uncertainty model was considered. Similar percentage increases for the whole parameter space with respect the case when no $\boldsymbol{n}(\boldsymbol{z})$ uncertainty model can found Table~\ref{tab: constraints}. Comparing these constraint with respects those of the traditional shifts \& stretches models shows that the PCA leads to a $5$ per cent wider $S_{\rm 8}$ constraint. Interestingly, the GP model yields a very similar level of uncertainty of the $S_{\rm 8}$ to that of the  shifts \& stretches model while yielding narrower $\Omega_{\rm m}$ and $\sigma_{\rm 8}$.

In order to better understand the origin of these discrepancies, we look into the interplay of the figure of merit parameter, $S_{\rm 8}$, with the parameters of the different galaxy distribution uncertainty models. We find that the different constraints on $S_{\rm 8}$ are linked to different galaxy distribution uncertainty models cutting the degeneracy with the galaxy bias parameters, $b_{\rm g}^i$, at different ranges. In Fig. \ref{fig: bias} we plot the 1D and 2D marginal posteriors of the parameters. Given the dimensionality of the parameter space, we plot the mean galaxy bias parameter, $\langle b_{\rm g} \rangle = \frac{1}{n_{\rm lenses}}\sum_i^{n_{\rm lenses}} b_{\rm g}^i$, across the lens tomographic bins. In this plot we can observe how the negative correlation between the two parameters leads to different $S_{\rm 8}$ constraints when the different galaxy redshift distribution models cut the degeneracy at different ranges.

We also investigate the redshift dependency of the interaction between the different galaxy distribution uncertainty models and galaxy bias parameters. In Fig. \ref{fig: bias_z} we show the evolution of the constraints on the galaxy bias parameters from each galaxy distribution uncertainty model as function of the mean redshift of their associated tomographic bin. We see that, as redshift increases, the relative difference between the bias constraints of each $\boldsymbol{n}(\boldsymbol{z})$ uncertainty model increases, meaning that going beyond the shifts \& stretches model has more impact for higher redshift bins where the uncertainty is the largest. Moreover, we also observe that the PCA model results in the higher standard deviation for the associated galaxy bias parameters. Notably, the GP model yields narrower galaxy bias constraints than the simples shift model for all the tomographic bins except the last. This suggests that the larger degradation of the $S_{\rm 8}$ constraint caused by the PCA and GP models is due to their impact on the highest redshift bin. 

\begin{figure*} 
\centering
\begin{subfigure}{0.47\textwidth}
  \centering
  \includegraphics[width=\linewidth]{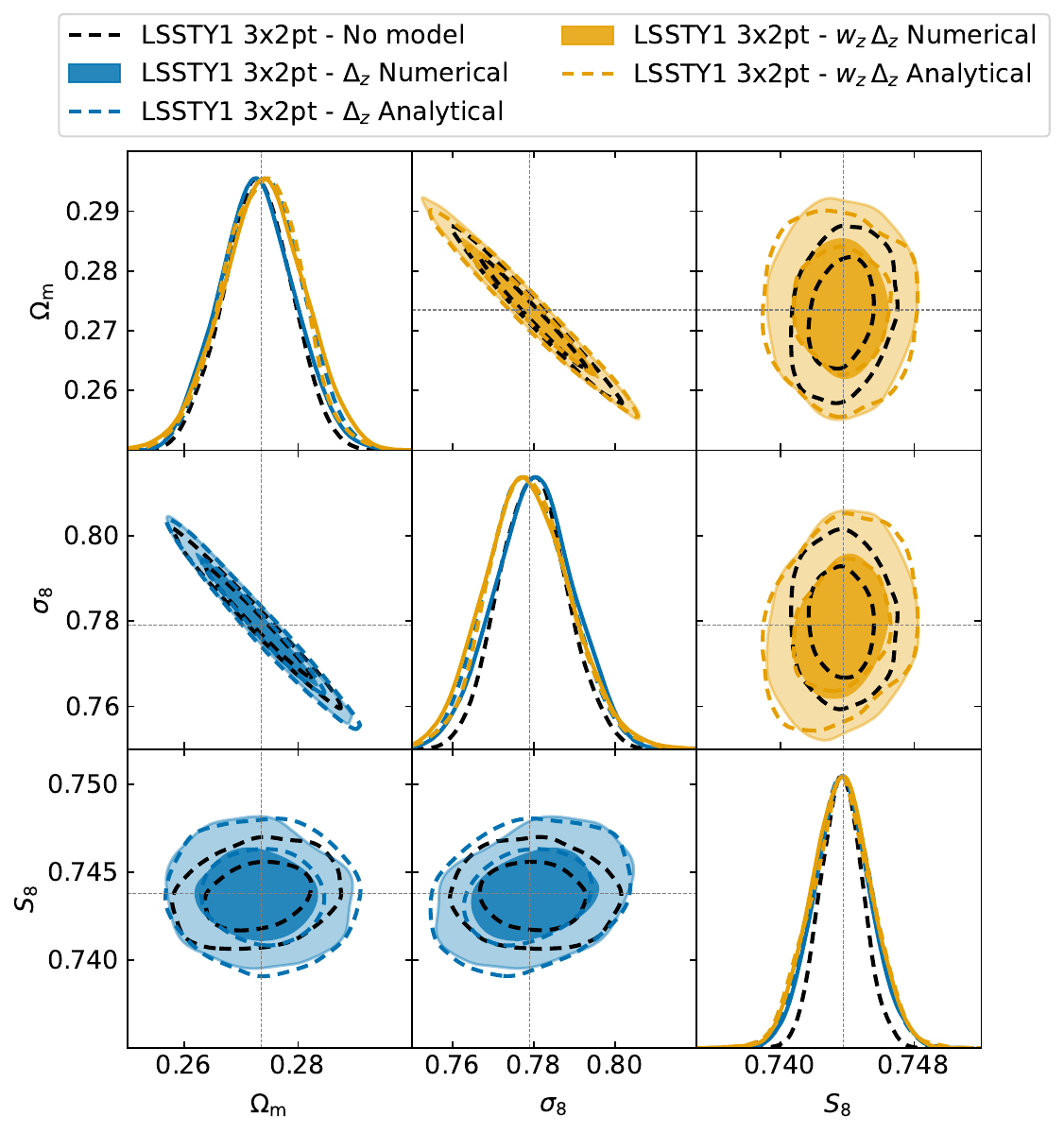}
\end{subfigure}%
\begin{subfigure}{0.47\textwidth}
  \centering
  \includegraphics[width=\linewidth]{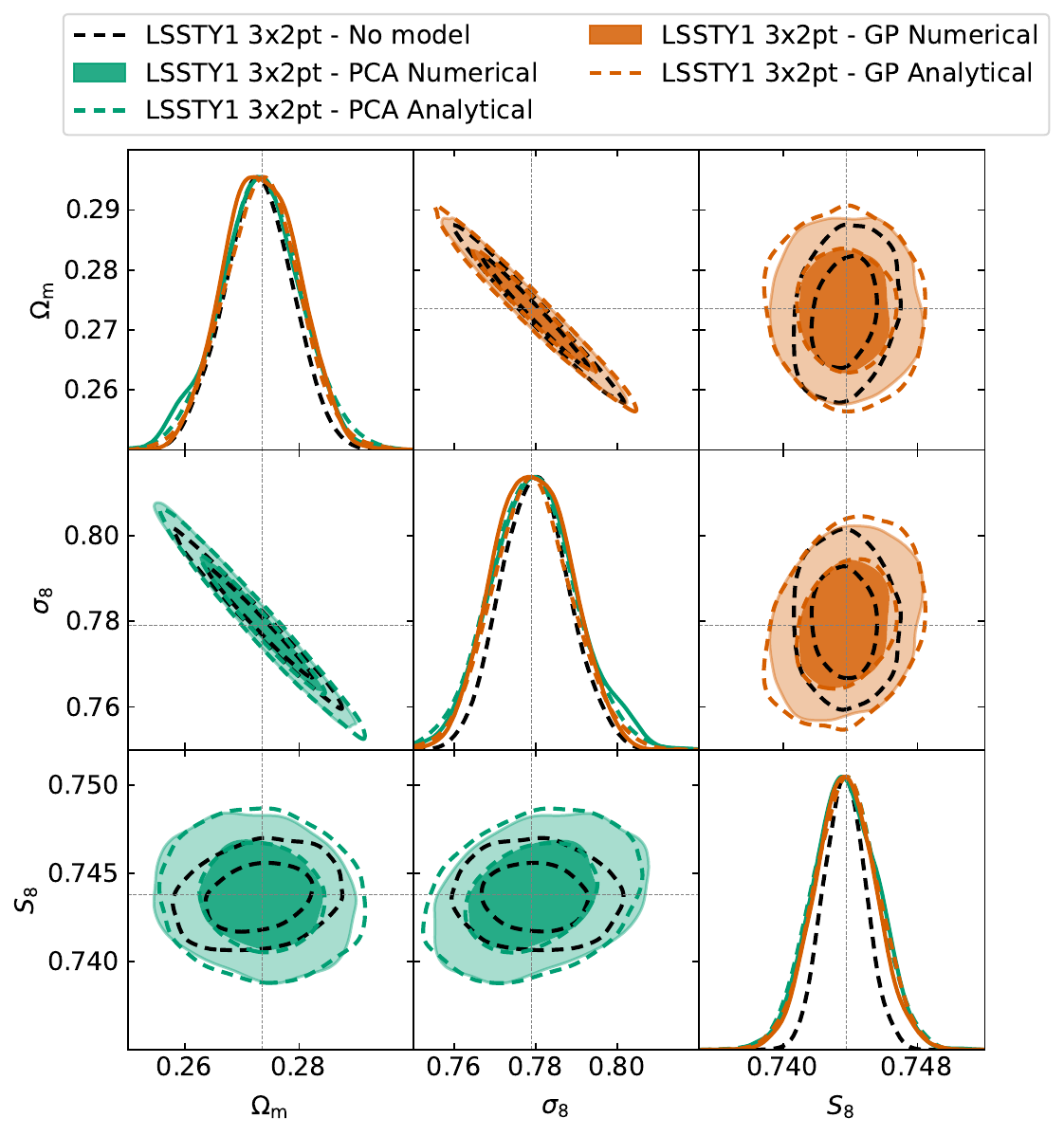}
\end{subfigure}
\caption{The 1 \& 2D marginalised constraints for the cosmological parameters $\Omega_{\rm m}$, $\sigma_{\rm 8}$ and $S_{\rm 8}$, accounting for $\boldsymbol{n}(\boldsymbol{z})$ uncertainties using different models and different marginalisations techniques. In each sub-panel three different contours are compared. First, a set of black dashed contours show the constraints obtained when the parameters of the associated model were kept fixed. Second, the two sets of coloured contours show the associated constraints when the parameters of the models were marginalised numerically and analytically.} \label{fig: ana_vs_num}
\end{figure*}

Having shown that the choice of galaxy distribution uncertainty model has a meaningful impact in the obtained cosmological constraints, we now investigate the potential biases incurred by each $\boldsymbol{n}(\boldsymbol{z})$ uncertainty model. We do so by evaluating our theory code at a fixed cosmology for each $\boldsymbol{n}(\boldsymbol{z})$ sample in the calibration ensemble. This generates generate an ensemble of fake data vectors based on the original redshift distributions. We then derive the parameter values of each  $\boldsymbol{n}(\boldsymbol{z})$ uncertainty model from each $\boldsymbol{n}(\boldsymbol{z})$ sample in the calibration ensemble. Assuming the same fixed cosmology, we evaluate our theory code at the mean $\boldsymbol{n}(\boldsymbol{z})$ of the calibration ensemble modified by each of the models given the parameter values associated with each sample. Finally, we compare the reference ensemble of data vectors with the ensemble generated by each model using a $\chi^2$ measure given by the covariance of our data. More formally, we compute: 
\begin{multline} \label{eq: bias_chi2}
    \chi^{2(i)} = [t(\boldsymbol{\theta_0}, n^{(i)}_{\rm true}(\boldsymbol{z}))-t(\boldsymbol{\theta_0}, n^{(i)}_{\rm model}(\boldsymbol{z}))]^T  \\ 
    \textsf{C}^{-1}[t(\boldsymbol{\theta_0}, n^{(i)}_{\rm true}(\boldsymbol{z}))-t(\boldsymbol{\theta_0}, n^{(i)}_{\rm model}(\boldsymbol{z}))] \,
\end{multline}
where $i$ denotes the ith sample in the $\boldsymbol{n}(\boldsymbol{z})$ calibration ensemble. Thus Eq. \ref{eq: bias_chi2} allows us to measure the error incurred by assuming a given $\boldsymbol{n}(\boldsymbol{z})$ model or, in other words, how much cosmological information is lost in the dimensionality reduction exercise. 

In Fig. \ref{fig: xi2_bias} we plot the respective $\tilde{\chi} \equiv \chi^2/(\rm dof)$ distributions for each model. Our results show that, of all models considered, the PCA model incurs by far the least model error yielding $\langle \tilde{\chi}^2_{\rm PCA}\rangle_{\rm \boldsymbol{n}(\boldsymbol{z})} = 0.005$, where $\langle \tilde{\chi}^2\rangle_{\rm \boldsymbol{n}(\boldsymbol{z})}$ represents the mean reduced $\chi^2$ over all the $\boldsymbol{n}(\boldsymbol{z})$ samples in the calibration ensemble. For comparison, the shifts model yields $\langle \tilde{\chi}^2_{\rm \Delta_{\rm z}}\rangle_{\rm \boldsymbol{n}(\boldsymbol{z})} = 0.031$, followed the by shifts \& stretches model with $\langle \tilde{\chi}^2_{\rm w_{\rm z} \Delta_{\rm z}}\rangle_{\rm \boldsymbol{n}(\boldsymbol{z})} = 0.025$ and the GP model with $\langle \tilde{\chi}^2_{\rm GP}\rangle_{\rm \boldsymbol{n}(\boldsymbol{z})} = 0.019$. We also study the case where no $\boldsymbol{n}(\boldsymbol{z})$ model is considered to establish a baseline, yielding $\langle \tilde{\chi}^2_{\rm no\, model}\rangle_{\rm \boldsymbol{n}(\boldsymbol{z})} = 0.16$. Note that since all other parameters except the $\boldsymbol{n}(\boldsymbol{z})$ uncertainty parameters are identical and the data vectors are noiseless, a perfect $\boldsymbol{n}(\boldsymbol{z})$ uncertainty model should be able to yield $\chi^2=0$ for every sample of the calibration ensemble.

In order to quantify the impact on the cosmological parameters of the previously computed goodness of fit distributions we consider linear perturbation on the cosmological parameters inside Eq. \ref{eq: bias_chi2} such that:
\begin{equation} \label{eq: theory_linear}
    t(\boldsymbol{\theta}, n^{(i)}_{\rm model}(z)) \approx  t(\boldsymbol{\theta_0}, n^{(i)}_{\rm model}(\boldsymbol{z})) + \textsf{J} \delta \boldsymbol{\theta}^{(i)} \,
\end{equation}
where $\textsf{J}$ is the Jacobian of the theory with respect to $\boldsymbol{\theta}$. By plugging Eq. \ref{eq: theory_linear} inside Eq. \ref{eq: bias_chi2} and optimising for $\delta \boldsymbol{\theta}$ we can obtain the bias on the cosmological parameters induced purely by the $\boldsymbol{n}(\boldsymbol{z})$ modelling error for each $\boldsymbol{n}(\boldsymbol{z})$ sample. Then, we can easily compute the mean bias over the ensemble of $\boldsymbol{n}(\boldsymbol{z})$ samples. In particular, we focus on the impact in the parameters of interest $(\Omega_{\rm m}, \sigma_{\rm 8})$. Given the linear nature of our approximation, we abstain from deriving biases on the $S_{\rm 8}$ parameter since it is a non-linear transformation of the previous parameters. In Fig. \ref{fig: bias_variance} we presented the obtained biases in the form of square root bias over standard deviation as a percentage. This metric is equivalent to considering "how many $\sigma$'s" off the truth an estimate on each parameter would be.

Our results show that that the shifts model leads to $0.16\sigma$ and $0.18\sigma$ bias on the parameters $(\Omega_{\rm m}$ and $\sigma_{\rm 8})$. Adding a stretch parameter reduces the biases to roughly $0.11\sigma$ on both parameters. In terms of the higher dimensional models, the GP model reduces the bias in $\sigma_{\rm 8}$ to $0.08\sigma$ but results in a higher $0.13\sigma$ bias on  $\Omega_{\rm m}$. On the other hand, the PCA model significantly outperforms all other model yielding a $0.07\sigma$ and $0.06 \sigma$ bias on $(\Omega_{\rm m}$ and $\sigma_{\rm 8})$ respectively. This is slightly over half the bias of the popular shifts \& stretches model. We note that the PCA model achieves this reduction in bias over variance mainly by reducing the bias in the parameters as shown by Fig. \ref{fig: bias}. While the PCA model also yields the largest variance (See Tab. \ref{tab: constraints}), it only degrades the $\Omega_{\rm m}$ and $\sigma_{\rm 8}$ constraints by less than 5 per cent with respect the shifts \& stretches model. Finally, it is important to bare in mind that this the bias is solely due to the modelling of $\boldsymbol{n}(\boldsymbol{z})$ uncertainty. 

Thus, we conclude that the PCA model strikes the best balance between bias and degradation of the constraint of the models considered in this work. Of course, the PCA model achieves this at the price of more than doubling the number of parameters per tomographic bin, significantly raising the cost of the inference. This begs the question whether there is a better way of accounting for these uncertainties in a fast and representative way.

\subsection{Validity of analytical marginalisation}

Studying the validity of analytical marginalisation of galaxy redshift distribution uncertainties is paramount due to the large proportion the associated parameters will represent of the total numbers of parameters that Stage-IV surveys with large numbers of tomographic bins will have to consider. As described in Section~\ref{sect: Laplace}, we use the Laplace approximation to account for the parameters of each of the galaxy distribution uncertainty models described in Section \ref{Sect: Redshift Uncertainty}. This is effectively done by running MCMC's where the $\boldsymbol{n}(\boldsymbol{z})$ uncertainty parameters are kept fixed but an additional term is added to the covariance of the data that accounts for their impact. 

We show our results in Fig. \ref{fig: ana_vs_num}, which shows the 1D \& 2D constraints for the cosmological parameters $\Omega_{\rm m}$, $\sigma_{\rm 8}$, and $S_{\rm 8}$ obtained when considering the four galaxy distribution uncertainty models studied in this work both analytically and numerically. We observe excellent consistency between the analytical and numerical constraints for all the models considered. The maximum discrepancy between the numerical and analytical constraints in the figure of merit we observe is approximately $5$ per cent, finding that the analytical constraints systematically overestimate the uncertainty in the figure of merit. In Fig. \ref{fig: S8_comp} we can observe how analytical marginalisation leads to 1D marginals for the $S_{\rm 8}$ parameter up to $5$ per cent wider than those from full numerical marginalisation for all the $\boldsymbol{n}(\boldsymbol{z})$ uncertainty models considered in this work. This means that the Gaussian assumption made by the Laplace approximation on the posterior of the parameters of the galaxy distribution uncertainty models is a conservative assumption that does not risk underestimating our confidence in the figure of merit. This a generic feature of the Laplace approximation since the prior is always expected to contain more volume than the posterior. Studies that require better precision on their estimates of the variance of their posteriors will thus have to rely on numerical methods.  These results are consistent with \citet{Boryana23} who performed a similar comparison between analytical versus numerical marginalisation in the context of a 3$\times$2-pt analyses of DESY1 data. Similarly to the previous section, we report the percentage increase in the standard deviation of parameter in the model with respect the case when no $\boldsymbol{n}(\boldsymbol{z})$ uncertainty model is consider in Table~\ref{tab: constraints}. It is worth noting that, while analytical marginalization generally leads to broader constraints, we find that the Laplace approximation yields tighter constraints for the shifts \& stretches model for the cosmological parameters $\Omega_{\rm m}$ and $\sigma_{\rm 8}$ as well as for the associated linear galaxy bias parameters. This indicates that the Laplace approximation doesn't properly capture the non-Gaussian degeneracy between the stretch parameters and the galaxy bias parameter, leading to artificially narrower constraints. While these discrepancies might appear large compared to the standard deviation of the posterior when no photometric uncertainties are included, it is important to stress that relative to numerical shifts \& stretches posterior they are, at worst, of order $7$ per cent.

Finally we are interested in studying how much time is saved by performing the marginalisation over the different photometric models analytically. In order to do so, we compare the Effective Sample Size (ESS) per second of the numerical and analytical chains. The ESS measures the number of uncorrelated samples in a Monte Carlo process. One can imagine that if the process was truly Markovian each sample would be completely uncorrelated from the previous one. However, different inference algorithms introduce an effective correlation length. Thus one can think of the ESS as the total number of samples divided by the correlation length. More formally, the ESS can be estimated as:
\begin{equation} \label{eq: ESS}
    {\rm{ESS}} \equiv \frac{N}{1+2\sum_{t=0}^\infty \rho_t} \, ,
\end{equation}
where $N$ is the number of samples in the chain and $\rho_t$ is the auto-correlation of the MCMC at a lag $t$ \citep{Carter&Khon}.

Once the ESS of a chain has been computed one can look at the ESS per second by dividing the ESS by the time taken to compute the chain. Since the convergence of a chain depends on the ESS, not the total number of samples, the ESS per second effectively measures how quickly a chain converges. By looking at the ratio of ESS per second between the analytical and numerical chains we can thus measure the impact of analytical marginalisation on the convergence rate of the chains. Note that this metric is unit-less and hardware agnostic since it does not depend on the likelihood evaluation time. We first compare the analytical chains to the reference chain where the galaxy redshift distribution uncertainties are not taken into account. As expected, our study shows that analytical marginalisation converges at the same rate when the uncertainties are not included in the analysis. We then turn our attention to the comparison between analytical and numerical marginalisation. Our results show that analytical marginalisation results in speed-ups of a factor of $\sim 5$ for the lower-dimensional models. For the higher-dimensional models, PCAs and GPs, we find speed-ups of a factor $\sim 25$. Thus, analytical marginalisation can deliver speed-ups by more than an order of magnitude at the expense of $5$ per cent errors on the variance of the derived posteriors. This stresses the importance of developing frameworks to make analytical marginalisation seamlessly implementable for Stage-IV analyses.

\section{Conclusions} \label{Sect: Conclusions}
\begin{figure}
    \centering
    \includegraphics[width=\linewidth]{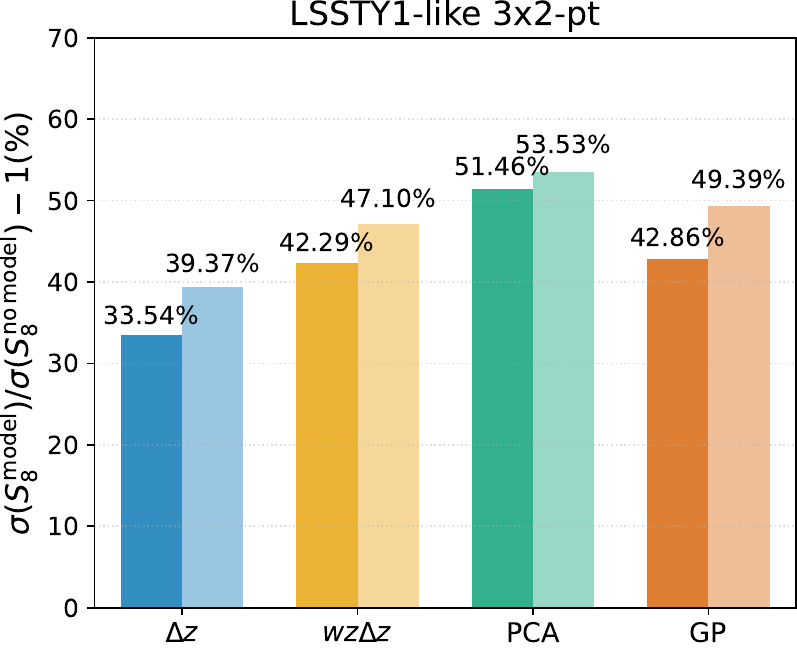}
    \caption{A comparison between the 1D marginal $S_{\rm 8}$ distributions obtained when considering different galaxy redshift distribution uncertainty models with respect the case where no model is considered. We show constraints for when a shift model (blue), a shift \& stretch model (yellow), a PCA model (green) and a GP model (orange). Full opacity bars correspond to numerical constraints while half opacity bars correspond to the equivalent analytical constraints.} \label{fig: S8_comp}
\end{figure}
In this work we studied the impact of different galaxy redshift distribution uncertainty models in the cosmological constraints of a 3$\times$2-pt analysis of mock Vera C. Rubin's observatory Legacy Survey of Space and Time Year 1 data, here onwards referred to as LSSTY1-like. In order to do so, we made use of the calibration ensemble of galaxy redshift distributions, $\boldsymbol{n}(\boldsymbol{z})$, by \citet{Yunhao2} designed to contain the expected level of both systematic and statistical uncertainty of LSSTY1 data. We then developed a new DESC Python package, \texttt{nz\_prior}, to turn this ensemble of possible $\boldsymbol{n}(\boldsymbol{z})$'s into calibrated priors for four galaxy distribution uncertainty models, namely: shifts, shifts \& stretches, principal components analysis (PCA),  and a Gaussian process (GP). 

Making use of the calibrated priors, we ran MCMC chains for each $\boldsymbol{n}(\boldsymbol{z})$ uncertainty model on the generated LSSTY1-like synthetic data vector with an appropriate Gaussian covariance. Due to the large dimensionality of some of the galaxy distribution uncertainty models, especially the PCA and GP models which add five extra parameters per tomographic bin, we used the self-tuning Hamiltonian Monte Carlo sampling scheme NUTS. Motivated by \citet{Ruiz_Zapatero_nzs}, we also studied the viability of marginalising over the parameters of those galaxy distribution uncertainty models analytically using the Laplace approximation. Finally, we investigated the obtained posteriors on the cosmological parameters to answer two questions: 
\\
\textbf{What the galaxy redshift distribution uncertainty model should be used in upcoming 3$\times$2-pt analyses? \\ PCA}: Leveraging our calibration ensemble of galaxy redshift distributions, we computed the distance between the angular power spectra obtained from the actual and the modelled $\boldsymbol{n}(\boldsymbol{z})$ for each sample in the ensemble. We then translated the observed modelling error into bias estimates on the cosmological parameters $\Omega_{\rm m}$ and $\sigma_{\rm 8}$ by linearising the theory prediction. Finally, we combined this measure of bias with the variance of the MCMC chains of each model to quantify their respective bias over variance ratio. Our results showed that, for a LSSTY1-like 3$\times$2-pt analysis, the PCA model incurs roughly half the amount of bias over variance in both parameters than the popular shift \& stretch model as shown Fig. \ref{fig: bias_variance}.  Moreover, our results show that, despite the significantly larger number of parameters, using five principal components per tomographic to model the uncertainty in the galaxy distributions only results in $5$ per cent wider constraints in the $S_{\rm 8}$ parameter than the traditional shift and stretch model (see Fig. \ref{fig: S8_comp}). Thus, the PCA model provides meaningfully less biased contours at the expense of 5 per cent more variance than the common shift and stretch model.\\
\\
\textbf{Can we marginalise over said model analytically? \\ Yes}: our results show that our analytical marginalisation scheme based on the Laplace approximation inflates errors up to $5$ per cent on the standard deviation of $S_{\rm 8}$ while speeding up the inference by a factor between 5 and 25 times depending on the dimensionality of the original model. The consistency between analytical and numerical $S_{\rm 8}$ constraints can be seen by comparing the dark and light bars in Fig. \ref{fig: S8_comp}. While this error is not negligible, we argue that, given the simplicity of the scheme and its potential to speed up the inference of heavy models by more than order of magnitude, the Laplace approximation should become a common-place tool for preliminary analyses as well as other use cases where high precision is not needed. These results are consistent with similar studies performed on 3$\times$2-pt analyses of DESY1 data for lower-dimensional models \citep{Boryana23}.\\
\\

Given the results shown in this paper, we see a number of potential avenues for future work. First and foremost, the results of this work show that upcoming analyses of LSSTY1 data should ready themselves to consider higher-dimensional $\boldsymbol{n}(\boldsymbol{z})$ uncertainty models, such as the PCA model we advocate for in this work, to avoid potential biases in their cosmological parameters. In order to do so effectively, auto-differentiable frameworks capable of accelerating the speed of the cosmological inference will be needed to efficiently test analysis choices as well as obtain the final constraints. Efforts are being undertaken towards achieving this goal. For example, the DESC collaboration is in the process of developing a \texttt{JAX} version of their theory code \texttt{CCL} \citep{ccl}, called \texttt{JAXCCL}, based on the \texttt{JAXCOSMO} library \citep{JAX-COSMO}. More effort will be needed to ready these libraries for the official analyses of the data.

Second, it is becoming increasingly clear that analytical marginalisation schemes are both reliable and fast methods to account for prior-dominated systematics. Particularly, in this work we showed the Laplace approximation is a safe way to include the impact of higher-dimensional $\boldsymbol{n}(\boldsymbol{z})$ uncertainty models at no additional computational cost. However, there is still no cosmological inference framework that would allow scientists to seamlessly implement analytical marginalisation over a subset of their parameter space. Thus, more efforts should be dedicated to creating such a tool. 

Third, in this work we proposed a data-driven methodology to quantify and propagate  uncertainties in the redshift distribution of galaxies based on generating an ensemble of possible $\boldsymbol{n}(\boldsymbol{z})$ distributions given the uncertainty in the photometric properties of each source in the catalogue. This ensemble was then fed to \texttt{nz\_prior} to calibrate different uncertainty models that can be plugged into a likelihood framework to propagate this uncertainty to cosmological constraints. Likelihood-free inference is a framework to perform statistical inference learning the posterior distribution of the cosmological parameters without the need to explicitly write a likelihood, based on the ability to simulate data realizations \citep{SBI, 2025MNRAS.536.1303J, 2024PhRvD.109f3534G, 2025PhRvD.111f3504G, SBI_HSC, SBI_KIDS}. Thus, the realizations of the $\boldsymbol{n}(\boldsymbol{z})$ used in this work to calibrate the priors of the different models could be used to directly learn the posterior within a likelihood-free framework. This approach would pose several advantages over the current methodology. First, there would be no need to choose an uncertainty model since it is only used within the likelihood. Second, marginalising over these sources of uncertainty, an expensive process that requires gradient-based inference methods within a likelihood framework, becomes trivial in likelihood-free inference. 

Finally, it is important to stress that the conclusions of this work should be understood in the context of a LSSTY1-like analysis and that extrapolating them to other scenarios should be done with care. We expect our results to be directly applicable to similar upcoming analyses to LSSTY1 such as Euclid DR1 \citep{Euclid_overview}. In order to forecast the importance of these results into the future we have to understand the two main factors that will determine the importance of $\boldsymbol{n}(\boldsymbol{z})$ uncertainty model. On the one hand, as the data precision increases the sensitivity to the galaxy redshift distribution will also increase, raising the importance of going beyond traditional shifts \& stretches $\boldsymbol{n}(\boldsymbol{z})$ uncertainty models. However, better calibration of the $\boldsymbol{n}(\boldsymbol{z})$ with smaller uncertainties will reduce the importance of the model used to capture them. Thus, future analyses will have to consider the interplay of these forces when assessing what $\boldsymbol{n}(\boldsymbol{z})$ uncertainty model to use. Moreover, detailed aspects such as the number of principal components needed or the accuracy of the analytical approximation are particular to the error budget and $\boldsymbol{n}(\boldsymbol{z})$ calibration considered in this work.

\section*{Software Availability}
\texttt{nz\_prior} is a DESC publicly available library that can found at: https://github.com/LSSTDESC/nz\_prior.

\section*{Contribution statement}
JRZ: Science and code implementation lead.
QH: Science co-lead.
YHZ: data provider.
BJ: Science advisor.
JZ: Science advisor.
IH: Reviewer.
CGG: DESC Builder, developed TJPCOV and its interface with TXPipe.
AM: DESC Builder, contributed to early project conceptualization and use of qp.
BS: Reviewer.

\section*{Acknowledgements}
JRZ would like to thank the Oxford BIPAC group for kindly lending the computing resources that made this project possible as well as their frequent and fruitful discussion. JRZ would also like to thank Edd Edmondson for supporting the Hypatia cluster. 

JRZ, QH, and BJ acknowledge support by the ERC-selected UKRI Frontier Research Grant EP/Y03015X/1 and by STFC grant ST/W001721/1.
YHZ and JZ are supported by STFC funding for UK participation in LSST, through grant ST/X001334/1. This paper has undergone internal review in the LSST Dark Energy Science Collaboration. The authors would like to thank the internal reviewers Benjamin Stölzner and Ian Harrison for their insightful and thoughtful comments. The DESC acknowledges ongoing support from the Institut National de Physique Nucl\'eaire et de Physique des Particules in France; the Science \& Technology Facilities Council in the United Kingdom; and the Department of Energy and the LSST Discovery Alliance in the United States.  DESC uses resources of the IN2P3 Computing Center (CC-IN2P3--Lyon/Villeurbanne - France) funded by the  Centre National de la Recherche Scientifique; the National Energy Research Scientific Computing Center, a DOE Office of Science User Facility supported by the Office of Science of the U.S.\ Department of Energy under Contract No.\ DE-AC02-05CH11231; STFC DiRAC HPC Facilities, funded by UK BEIS National E-infrastructure capital grants; and the UK particle physics grid, supported by the GridPP Collaboration.  This work was performed in part under DOE Contract DE-AC02-76SF00515.

For the purpose of Open Access, the authors have applied a CC BY public copyright licence to any Author Accepted Manuscript version arising from this submission.

We made extensive use of the {\tt numpy} \citep{van2011numpy}, {\tt scipy} \citep{scipy_virtnamen}, {\tt astropy} \citep{astropy_2013, astropy_2018}, {\tt healpy} \citep{Zonca2019}, {\tt GetDist} \citep{getdist}, and {\tt matplotlib} \citep{Hunter_2007} python packages. We also make use of the \texttt{Julia} packages {\tt ForwardDiff.jl} \citep{ForwardDiff},  {\tt AdvancedHMC.jl}  \citep{AHMC} and {\tt Turing.jl} \citep{ge2018t}.

\bibliographystyle{mnras}
\bibliography{main}

\appendix
\section{Full Constraints}
\begin{table}
   \centering
   \renewcommand{\arraystretch}{1.5}
   \begin{tabular}{|c|c|c|c|c|}
   \hline
   \multicolumn{5}{|c|}{LSSTY1-like 3$\times$2-pt Analysis} \\
   \hline
    Numerical  & \multirow{2}{*}{$\Delta z$} & \multirow{2}{*}{$wz\Delta z$} & \multirow{2}{*}{PCA} & \multirow{2}{*}{GP} \\
   Analytical & & & & \\
   \hline
   \hline
\multirow{2}{*}{$\Omega_{\rm m}$} & 11.52\% & 23.33\% & 19.84\% & 8.13\% \\
                                  & 14.60\% & 13.89\% & 20.09\% & 12.43\% \\
\hline
\multirow{2}{*}{$\sigma_8$} & 17.82\% & 27.93\% & 28.50\% & 12.56\% \\
                            & 20.39\% & 20.24\% & 27.44\% & 18.81\% \\
\hline
\multirow{2}{*}{$S_8$} & 33.54\% & 42.29\% & 51.46\% & 42.86\% \\
                       & 39.37\% & 47.10\% & 53.53\% & 49.39\% \\
\hline
\multirow{2}{*}{$n_{\rm s}$} & 10.28\% & 12.46\% & 13.04\% & 8.28\% \\
                             & 14.37\% & 12.20\% & 13.95\% & 4.48\% \\
\hline
\multirow{2}{*}{$A_{\rm IA}$} & 39.36\% & 47.34\% & 45.45\% & 45.94\% \\
                              & 38.78\% & 41.36\% & 48.13\% & 46.21\% \\
\hline
\multirow{2}{*}{$b_{\rm g}^0$} & 16.28\% & 25.27\% & 27.98\% & 13.80\% \\
                               & 18.52\% & 19.73\% & 26.60\% & 19.50\% \\
\hline
\multirow{2}{*}{$b_{\rm g}^1$} & 18.55\% & 32.73\% & 31.52\% & 14.96\% \\
                               & 22.13\% & 27.80\% & 31.75\% & 21.02\% \\
\hline
\multirow{2}{*}{$b_{\rm g}^2$} & 22.55\% & 33.68\% & 34.40\% & 14.67\% \\
                               & 22.95\% & 28.64\% & 33.42\% & 21.90\% \\
\hline
\multirow{2}{*}{$b_{\rm g}^3$} & 23.81\% & 36.38\% & 39.16\% & 18.70\% \\
                               & 26.16\% & 31.42\% & 36.04\% & 25.15\% \\
\hline
\multirow{2}{*}{$b_{\rm g}^4$} & 20.89\% & 41.12\% & 64.10\% & 45.83\% \\
                               & 23.10\% & 33.59\% & 60.41\% & 48.03\% \\
\hline
\end{tabular}
\caption{Percentage increase in the standard deviation of the constraints on cosmological and nuisance parameters obtained in a 3$\times$2-pt analysis of LSSTY1-like data when considering different galaxy distribution uncertainty models with respect the case when no model is considered. In each cell we display two values, the upper value corresponds to the numerical constraint obtained using NUTS and the lower the analytical Laplace approximation. We exclude flat, prior dominated parameters, such as $\Omega_{\rm b}$ and $h$ since in these cases the standard deviation can be misleading. All MCMC's in this analysis were run until the Monte Carlo error in the standard deviation of the parameters was below 1 per cent the standard deviation of the parameter.}
\label{tab: constraints}
\end{table}

\bsp	
\label{lastpage}
\end{document}